\documentclass{emulateapj} 
\usepackage{morefloats}
\usepackage{amsmath}
\usepackage{verbatim}
\usepackage{color}
\usepackage{graphicx}
\usepackage[colorlinks,urlcolor=blue,citecolor=blue,linkcolor=blue]{hyperref}
\usepackage{floatrow}
\newcommand{\be}{\begin{eqnarray}}
\newcommand{\ee}{\end{eqnarray}}
\newcommand{\bi}{\begin{itemize}}
\newcommand{\ei}{\end{itemize}}

\begin{document}

\title{Non-Thermal Electron Acceleration in Low Mach Number Collisionless Shocks.\\
$\!$II. Firehose-Mediated Fermi Acceleration and its Dependence on Pre-Shock Conditions}
\author{
Xinyi Guo$^1$, Lorenzo Sironi$^{1,2}$ and Ramesh Narayan$^1$
}
\affil{$^1$ Harvard-Smithsonian Center for Astrophysics, 60 Garden St., Cambridge, MA 02138, USA \\$^2$ NASA Einstein Postdoctoral Fellow
}
\slugcomment{Accepted to ApJ on October 23, 2014}
\begin{abstract}
Electron acceleration to non-thermal energies is known to occur in low Mach number ($M_s\lesssim 5$) shocks in galaxy clusters and solar flares, 
but the electron acceleration mechanism remains poorly understood. 
Using two-dimensional (2D) particle-in-cell (PIC) plasma simulations, we showed in 
Paper I that electrons are efficiently accelerated in low Mach number ($M_s = 3$) quasi-perpendicular shocks via a Fermi-like process. The electrons bounce between the upstream region and the shock front, with each reflection at the shock resulting in energy gain via shock drift acceleration. The upstream scattering is provided by oblique magnetic waves, that are self-generated by the electrons escaping ahead of the shock.
In the present work, we employ additional 2D PIC simulations to address the nature of the upstream oblique waves.
We find that the waves are generated by the shock-reflected electrons via the firehose instability, 
which is driven by an anisotropy in the electron velocity distribution. We systematically explore how the efficiency of wave generation and of electron acceleration depend on the magnetic field obliquity, the flow magnetization (or equivalently, the plasma beta), and the upstream electron temperature. 
We find that the mechanism works for shocks with high plasma beta ($\gtrsim 20$) at nearly all magnetic field obliquities, and for electron temperatures in the range relevant for galaxy clusters.
Our findings offer a natural solution to the conflict between the bright radio synchrotron emission observed from the outskirts of galaxy clusters and the low electron acceleration efficiency usually expected in low Mach number shocks.
\end{abstract}

\section{Introduction}
There is considerable observational evidence that electrons are efficiently accelerated in low Mach number collisionless shocks in astrophysical sources. 
In particular, at the outskirts of galaxy clusters, 
where X-ray telescopes have unambiguously 
detected the existence of low Mach number shocks based on density and/or temperature jumps,
radio observations reveal synchrotron emission 
from relativistic electrons, presumably accelerated at the shock fronts
\citep[e.g.][]{Markevitch2007,Finoguenov2010,vanWeeren2010,Akamatsu2012,Bruggen2012,Feretti2012,Brunetti2014}.
However, the physics of the electron acceleration mechanism remains poorly understood.
This paper is the second in a series, focusing on the study of electron acceleration in low Mach number shocks by means of self-consistent PIC simulations. 

In the first paper of this series \citep[][Paper I hereafter]{Guo2014a}, we 
focused on the particle energy spectra and the acceleration mechanism in a reference PIC run with Mach number $M_s = 3$
 and a quasi-perpendicular magnetic field. 
We found that about $15\%$ of electrons are efficiently accelerated, forming a non-thermal power-law tail 
in the energy spectrum with a slope of $p\simeq2.4$. We identify the acceleration mechanism to be as follows.
A fraction of the incoming electrons are energized at the shock front via shock drift acceleration (SDA). The accelerated electrons are reflected back upstream by the mirror force of the shock-compressed magnetic field. 
In the upstream region, the interaction of these electrons with the incoming flow generates magnetic waves. In turn, the waves scatter some of the electrons propagating upstream back toward the shock, for further energization via SDA. Thus the self-generated waves allow for repeated cycles of shock drift acceleration, similar to a sustained Fermi-like process.  

In Paper I, we did not investigate the nature of the upstream waves, which are essential
for sustaining the Fermi-like process. We show in this work that the waves are
triggered by the shock-reflected electrons propagating upstream, via the electron firehose instability. In addition to clarifying the nature of the upstream waves, another goal of this paper is to explore the dependence of the 
efficiency of firehose-mediated electron acceleration on pre-shock conditions. 

The reference run in Paper I was set up to capture the physical environment at the Galactic Center along the trajectory of the G2 cloud \citep{Narayan2012,Olek2013}, 
where the plasma temperature $T$ is very high, reaching $k_B T\sim 100\ \rm{keV}$, 
where $k_B$ is the Boltzmann constant. 
In the intracluster medium (ICM), where low Mach number shocks
frequently occur due to mergers, the plasma temperature is lower, $k_B T \sim 1-10\ \rm{keV}$. 
The magnetic field pressure in the reference run was chosen to be a fraction $\sim5\%$ of the plasma thermal pressure, and the field was quasi-perpendicular to the shock direction of propagation (with obliquity $\theta_B=63^\circ$).
Observationally, the magnetic field strength and obliquity cannot be easily constrained, 
though we expect a range of strengths and obliquities. 
With this as motivation, we explore here the dependence of the electron acceleration mechanism on various pre-shock parameters.

The paper is organized as follows.
In Section \ref{sec:setup}, we describe the simulation setup and our choice of physical parameters. 
In Section \ref{sec:shockstructure}, we summarize the shock structure and the electron acceleration mechanism described for the reference run in Paper I.
In Section \ref{sec:SDA}, we investigate the dependence of the SDA injection process on the pre-shock conditions. 
In Section \ref{sec:waves}, we study in detail the nature of the upstream waves, which are essential for sustaining long-term acceleration of electrons. 
In Section \ref{sec:dependence}, we explore the dependence of the acceleration mechanism
on the upstream magnetic  obliquity, the flow magnetization (or equivalently, the plasma beta) and the electron temperature. 
We conclude with a summary and discussion of our findings in Section \ref{sec:discussion}.

\section{Simulation Setup and Parameter Choice}
\label{sec:setup}
We perform numerical simulations using the three-dimensional (3D) electromagnetic PIC code TRISTAN-MP \citep{Spitkovsky05}, which is a parallel version of the publicly available code TRISTAN \citep{Buneman1993} that has been optimized for studying collisionless shocks. 

The computational setup and numerical scheme are described in detail in Paper I. 
In brief, the shock is set up by reflecting an upstream electron-ion plasma off a conducting wall at the left boundary $(x=0)$ of the computational box. The upstream plasma is initialized as a Maxwell-J\"uttner distribution with the electron temperature $T_e$ equal to the ion temperature $T_i$, drifting with a bulk velocity $\vec{u}_0 = -u_0\hat{x}$. 
The interaction between the reflected stream and the incoming plasma causes a shock to form, which propagates along $+\hat{x}$ at a speed $u_{\rm sh}$. 
The relation between the upstream bulk flow velocity and the plasma temperature is parametrized by the simulation-frame Mach number 
\begin{equation}\label{eq:M}
M \equiv \frac{u_0}{c_s} = \frac{u_0}{\sqrt{2\Gamma k_B T_i/m_i}}~,
\end{equation} where $c_s$ is the sound speed in the upstream, $k_B$ is the Boltzmann constant, $\Gamma=5/3$ is the adiabatic index of the plasma, and $m_i$ 
is the ion mass.
The incoming plasma carries a uniform magnetic field $\vec{B}_0$, whose strength is parametrized by the magnetization
\begin{equation}
\sigma\equiv \frac{B_0^2/4\pi}{ \left(\gamma_0-1\right)n_0 m_ic^2 }~,
\end{equation}
where $\gamma_0 \equiv \left(1-u_0^2/c^2\right)^{-1/2}$ is the upstream bulk Lorentz factor 
and $n_0=n_i = n_e$ is the number density of the incoming plasma.
The magnetic field orientation with respect to the shock normal (aligned with $+\hat{x}$) is parameterized by the polar angle $\theta_{B}$ and the azimuthal angle $\varphi_{B}$, where $\varphi_{B}=0^\circ$ if the magnetic field lies in the $xy$ plane of the simulations.
The incoming plasma is initialized with zero electric field in its rest frame. Due to its bulk motion, the upstream plasma carries a motional electric field $\vec{E}_0 = -(\vec{u}_0/c)\times\vec{B}_0$ in the simulation frame.

In the literature, the Mach number $M_s$ is often defined as the ratio between the upstream flow velocity and the upstream sound speed in the shock rest frame (rather than in the downstream frame, as in Equation \eqref{eq:M}). In the limit of 
weakly magnetized shocks, the shock-frame Mach number $M_s$ is related to our simulation-frame Mach number $M$ through the implicit relation
\begin{equation}\label{eq:MsM}
M_s = M\frac{u_{\rm sh}^{\rm up}}{u_0}=M\left(1+\frac{1}{r\left(M_s\right)-1}\right),
\end{equation}
where $u_{\rm sh}^{\rm up}$ is the shock velocity in the upstream rest frame, equal to the upstream flow velocity in the shock rest frame, and 
\begin{equation}\label{eq:r}
r\left(M_s\right)= \frac{\Gamma+1}{\Gamma-1+2/M_s^2}
\end{equation}
is the Rankine-Hugoniot relation for the density jump from upstream to downstream.

For comparison with earlier work, where the magnetic field strength 
is sometimes parametrized by the Alfv\'enic Mach number  $M_{A}\equiv u_0/v_A$, where $v_A \equiv B_0/\sqrt{4\pi n m_i}$ is the Alfv\'en velocity, we remark that the relation between the magnetization and the Alfv\'enic Mach number is simply $M_A = \sqrt{2/\sigma }$. Alternatively, one could employ the plasma beta $\beta_{p} \equiv 8\pi nk_B\left(T_e + T_i\right)/B_0^2$, which is related to the magnetization as $\beta_p = 4/\left( \sigma\Gamma M^2\right)$, under the assumption of temperature equilibrium $T_e = T_i$.
We stress that in our simulations the upstream particles are
initialized with the physically-grounded Maxwell-J\"uttner distribution, instead of the so-called ``$\kappa$-distribution'' that was employed by, e.g., \cite{Park2013}.
The $\kappa$-distribution might be a realistic choice for shocks in solar flares. However, in most other astrophysical settings one expects the upstream particles to populate a Maxwellian distribution. By using a $\kappa$-distribution, which artificially 
boosts the high-energy component of the particle spectrum, one would unphysically overestimate the electron acceleration efficiency.

In Paper I, we performed simulations in both 2D and 3D computational domains. We found that most of the shock physics is well captured by 2D simulations in the $xy$ plane, if the magnetic field lies in the simulation plane, i.e., $\varphi_{B} = 0^\circ$. Therefore, to  explore a wide rage of parameter space with fixed computational resources, in this paper we only utilize 2D runs with in-plane fields. We stress that all three components of particle velocities and electromagnetic fields are tracked. As a result, the adiabatic index is $\Gamma=5/3$.

For accuracy and stability, PIC codes have to resolve the plasma oscillation frequency of the electrons 
\begin{equation}
\omega_{pe} = \sqrt{4\pi e^2n_0/m_e}~,
\end{equation} 
and the electron plasma skin depth $c/\omega_{pe}$, where $e$ and $m_e$ are the electron charge and mass. On the other hand, the shock structure is controlled by the ion Larmor radius 
\begin{equation}\label{eq:rLi}
r_{{\rm L},i}=\sqrt{\frac{2}{\sigma}}\sqrt{\frac{m_i}{m_e}}\ \frac{c}{\omega_{pe}}\gg \frac{c}{\omega_{pe}}~,
\end{equation} 
and the evolution of the shock occurs on a time scale given by the ion Larmor gyration period $\Omega_{ci}^{-1}=r_{{\rm L},i}u_0^{-1}\gg \omega_{pe}^{-1}$.
The need to resolve the electron scales, and at the same time to capture the shock evolution for many $\Omega_{ci}^{-1}$, is an enormous computational challenge, especially for the realistic mass ratio $m_i/m_e=1836$. We found in Paper I that simulations with two choices of the mass ratio, $m_i/m_e =100$ and $m_i/m_e = 400$, show consistent results, so that the shock physics can be confidently extrapolated to the realistic mass ratio $m_i/m_e=1836$, using the scalings presented in Appendix B of that paper.
We therefore employ a reduced mass ratio $m_i/m_e = 100$ for all the runs presented in this paper. 

We adopt a spatial resolution of $10$ computational cells per electron skin depth $c/\omega_{pe}$ and we use a time resolution of $dt = 0.045\ \omega_{pe}^{-1}$. 
Each cell is initialized with $32$ particles ($16$ per species).  
The transverse box size is fixed at $76\ c/\omega_{pe}$. We have performed convergence tests which show that 5 cells per electron skin depth can resolve the acceleration physics reasonably well, and  we have confirmed that simulations with a number of particles per cell up to $64$ and a transverse box size up to $256\ c/\omega_{pe}$ show essentially the same results. 

\begin{table}
\begin{center}
\vspace{-0.1in}
\begin{tabular}{cccccc}
\hline 
Run  & $T_{e}=T_{i}$ {[}K({\rm keV}){]} & $u_{0}/c$ & $\theta_{B}$ &  $\sigma$ & $\beta_p$ \tabularnewline
\hline 
\hline 
\texttt{reference} & $10^{9}(86)$ & $0.15$ & $63^{\circ}$ &  $0.03$ & $20$\tabularnewline
\texttt{theta13} & $10^{9}(86)$ & $0.15$ & $13^{\circ}$ & $0.03$ & $20$\tabularnewline
\texttt{theta23} & $10^{9}(86)$ & $0.15$ & $23^{\circ}$ & $0.03$ & $20$\tabularnewline
\texttt{theta33} & $10^{9}(86)$ & $0.15$ & $33^{\circ}$ & $0.03$ & $20$\tabularnewline
\texttt{theta43} & $10^{9}(86)$ & $0.15$ & $43^{\circ}$ & $0.03$ & $20$\tabularnewline
\texttt{theta53} & $10^{9}(86)$ & $0.15$ & $53^{\circ}$ & $0.03$ & $20$\tabularnewline
\texttt{theta68} & $10^{9}(86)$ & $0.15$ & $68^{\circ}$ & $0.03$ & $20$\tabularnewline
\texttt{theta73} & $10^{9}(86)$ & $0.15$ & $73^{\circ}$ & $0.03$ & $20$\tabularnewline
\texttt{theta80} & $10^{9}(86)$ & $0.15$ & $80^{\circ}$ & $0.03$ & $20$\tabularnewline
\texttt{sig1e-1\_43} & $10^{9}(86)$ & $0.15$ & $43^{\circ}$ & $0.1$ & $6$\tabularnewline
\texttt{sig1e-1\_53} & $10^{9}(86)$ & $0.15$ & $53^{\circ}$ & $0.1$ & $6$\tabularnewline
\texttt{sig1e-1\_63} & $10^{9}(86)$ & $0.15$ & $63^{\circ}$ & $0.1$ & $6$\tabularnewline
\texttt{sig1e-1\_73} & $10^{9}(86)$ & $0.15$ & $73^{\circ}$ & $0.1$ & $6$\tabularnewline
\texttt{sig1e-2\_43} & $10^{9}(86)$ & $0.15$ & $43^{\circ}$ & $0.01$ & $60$\tabularnewline
\texttt{sig1e-2\_53} & $10^{9}(86)$ & $0.15$ & $53^{\circ}$ & $0.01$ & $60$\tabularnewline
\texttt{sig1e-2\_63} & $10^{9}(86)$ & $0.15$ & $63^{\circ}$ & $0.01$ & $60$\tabularnewline
\texttt{sig1e-2\_73} & $10^{9}(86)$ & $0.15$ & $73^{\circ}$ & $0.01$ & $60$\tabularnewline
\texttt{sig3e-3\_63} & $10^{9}(86)$ & $0.15$ & $63^{\circ}$ & $0.003$ & $200$\tabularnewline
\texttt{Te1e7.5} & $ 10^{7.5}(2.7)$ & $0.027$ & $63^{\circ}$ & $0.03$ & $20$\tabularnewline
\texttt{Te1e8.0} & $10^{8}(8.6)$ & $0.047$ & $63^{\circ}$ &  $0.03$ & $20$\tabularnewline
\texttt{Te1e8.5} & $10^{8.5}(27)$ & $0.084$ & $63^{\circ}$ & $0.03$ & $20$\tabularnewline
\hline 
\end{tabular}
\caption{Upstream Parameters Used for the Shock Simulations}
\label{table:params}
\end{center}
\end{table}

We carry out several runs with various values of $T_e$, $\sigma$ and $\theta_B$, while keeping $M=2$ (corresponding to $M_s = 3$) fixed.
The choice of $M_s = 3$ is representative of the Mach numbers of merger shocks in galaxy clusters ($M_s \sim 1.5 - 5$) in cosmological simulations \citep[e.g.][]{Ryu2003} and also inferred from X-ray and radio observations \citep[e.g.][]{Akamatsu2012}. The effect of varying $M_s$ has been explored by means of PIC simulations in \cite{Narayan2012}, and we shall briefly comment on the dependence on Mach number in Section \ref{sec:dependence}. 
The upstream parameters of our runs are summarized in Table \ref{table:params}.
We vary $T_e$ from $10^{7.5}K$ to $10^9K$, which overlaps the typical temperature range of the ICM ($T\sim 10^7-10^8K$). 
The magnetization $\sigma$ varies from $0.003$ to $0.1$, corresponding to a plasma beta ranging from
$\beta_p = 200$ to  $\beta_p = 6$, which is well motivated, based on the typical number density in the ICM ($10^{-4}-10^{-2}\ {\rm cm}^{-3}$)
and on the magnetic field strength \citep[a few $\mu \rm G$, see e.g.][]{Brunetti2014}.
We vary the obliquity angle $\theta_B$ across a wide range (from $13^\circ$ up to $80^\circ$, i.e., from quasi-parallel to quasi-perpendicular shocks), as it is usually not constrained by observations.

\section{Shock Structure and Particle Acceleration}
\label{sec:shockstructure}
\begin{figure*}
\begin{center}
\includegraphics[height=0.27\textheight]{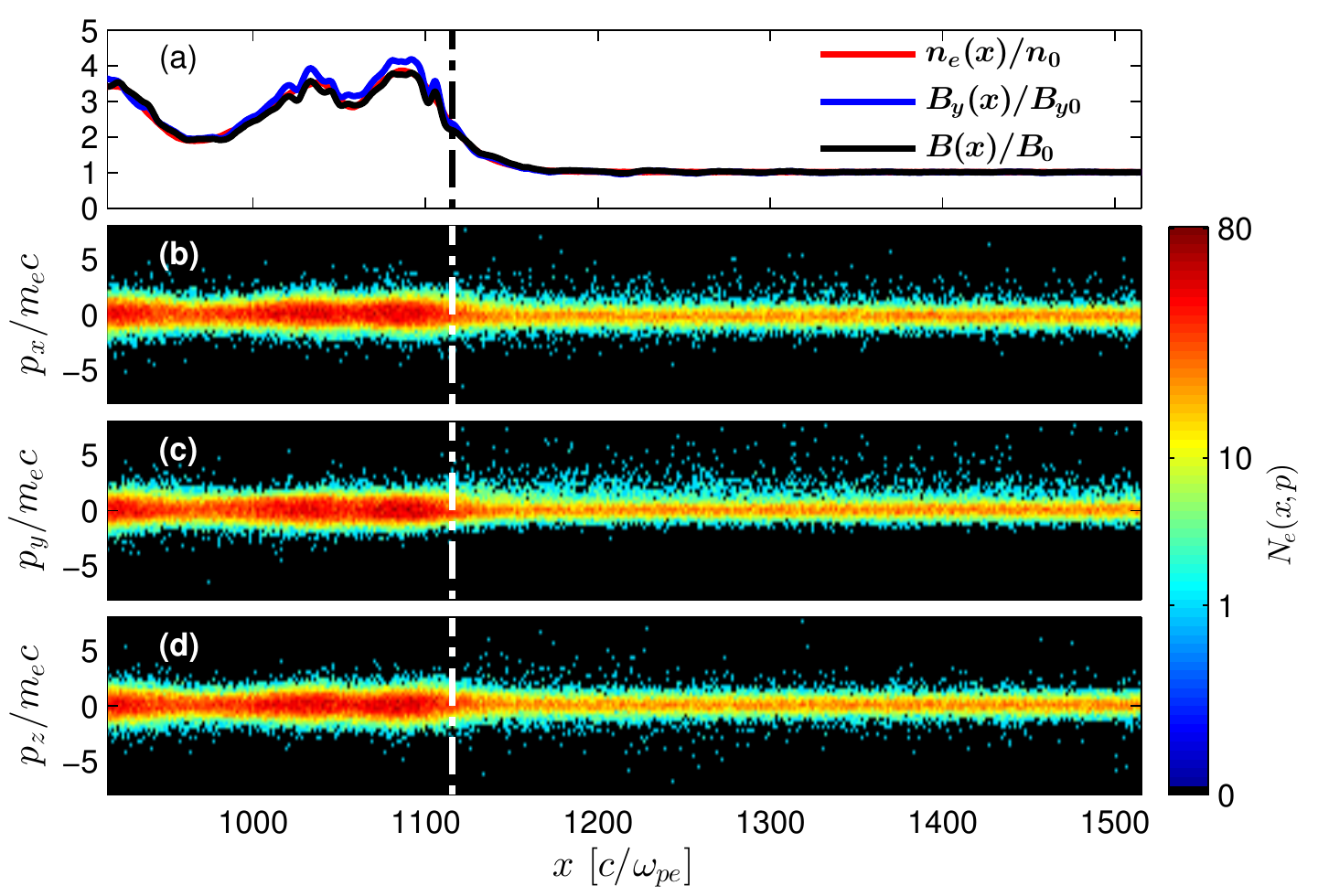}\includegraphics[height=0.27\textheight]{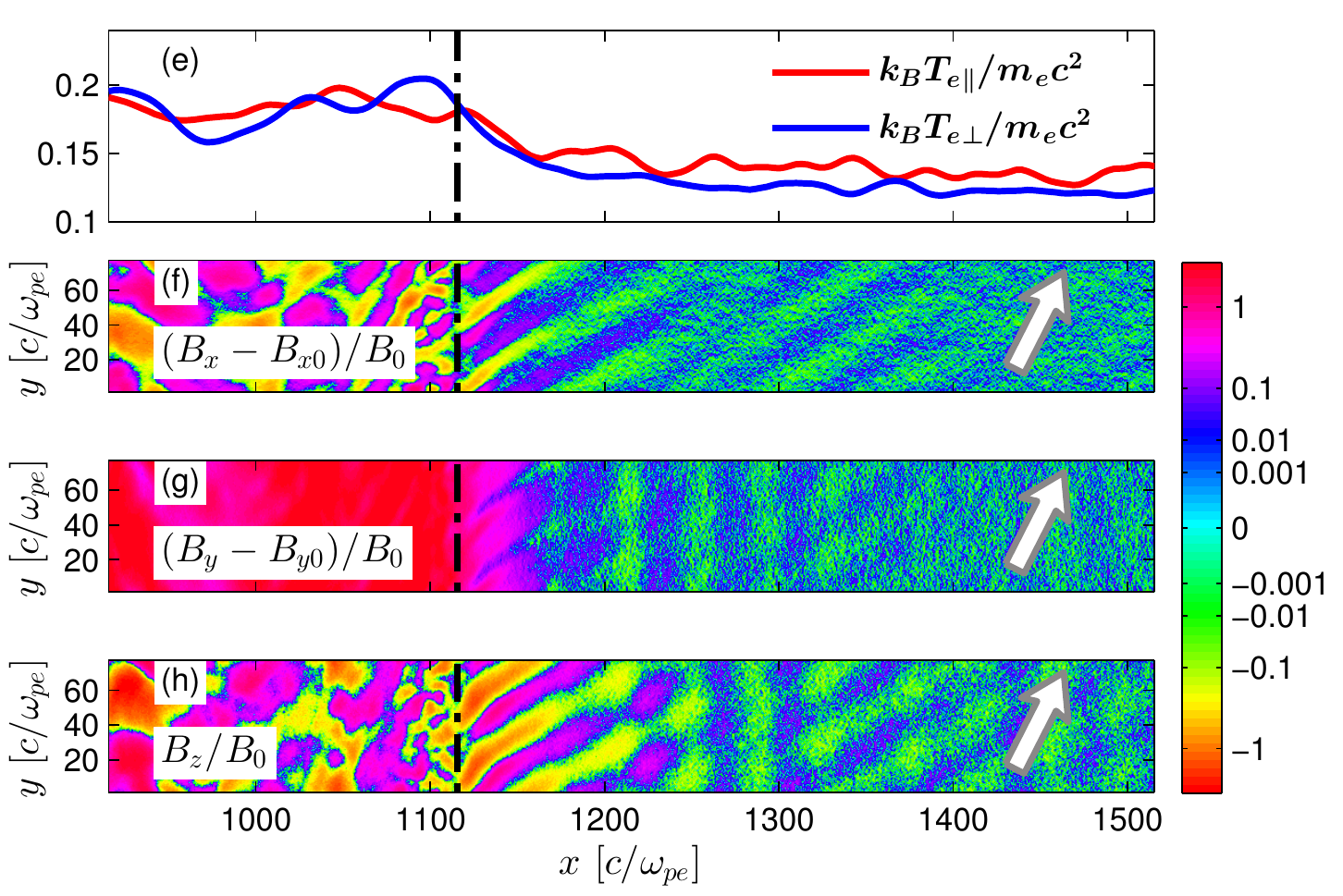}
\caption{
Shock structure of the \texttt{reference} run at time $\omega_{pe}t = 14625 \ (\Omega_{ci}t=26.9)$.
The shock is at $x\simeq1115\ c/\omega_{pe}$, as indicated by the vertical dot-dashed lines, 
and moves to the right. Downstream is to the left of the shock, and upstream to the right.
Panel (a) shows the ratios $n_e/n_0$ (red line), $B_y/B_{y0}$ (blue line) and $B/B_0$ (black line). 
Panels (b)-(d) show the electron momentum phase spaces
$p_x$-$x$, $p_y$-$x$, $p_z$-$x$, as a function of the longitudinal coordinate $x$. Panel (e) shows 
the electron temperatures parallel ($T_{e\parallel}$) and perpendicular ($T_{e\perp}$) to the magnetic field.  
Panels (f)-(h) show 2D plots of the magnetic field components in units of $B_0$, after subtracting the background field $\vec{B}_0$ (i.e., we show $(B_x-B_{x,0})/B_0$, $(B_y-B_{y,0})/B_0$ and $B_z/B_0$, respectively).
 The white arrows indicate the orientation of the upstream background magnetic field $\vec{B}_0$. Note that there are upstream waves in all three components 
of the magnetic field.}
\label{fig:shockstruct}
\end{center}
\end{figure*}

In Paper I, we analyzed the \texttt{reference} run (see Table \ref{table:params}) and showed
that electrons are efficiently 
accelerated, with the upstream electron 
energy spectrum developing a clear non-thermal component over time. 
Before exploring the parameter dependence 
of the electron acceleration efficiency, we first 
summarize the shock structure and the electron acceleration mechanism 
as inferred from the \texttt{reference} run in Paper I. 

Electrons are initially energized via shock drift acceleration (SDA) at the shock front.
At  the shock front (located at $x\simeq 1115\ c/\omega_{pe}$ 
in Fig. \ref{fig:shockstruct}), a fraction of the upstream thermal electrons get reflected by the mirror force of the shock-compressed magnetic field. The change in the total 
magnetic field strength $B/B_0$ (black curve in Fig. \ref{fig:shockstruct}(a)) is dominated by the compression of the perpendicular component $B_y$ (blue curve in Fig. \ref{fig:shockstruct}(a)), 
which is related via flux freezing to the density compression 
(red curve in Fig. \ref{fig:shockstruct}(a)). The plasma density increases 
at the shock by a factor of $\sim 4$, which is higher than the prediction ($\sim3$) of the Rankine-Hugoniot jump condition (see Equation \eqref{eq:r}). This density overshoot is a common feature of quasi-perpendicular shocks ($\theta_B\gtrsim 45^\circ$), both in the non-relativistic \citep[see e.g.][]{Hoshino2001,Treumann2009,Umeda2009,Matsumoto2012} and in the ultra-relativistic regime \citep[e.g.,][]{SS11,SSA13}. It is due to  Larmor gyration of the incoming ions in the compressed shock fields, which decelerates the upstream flow and therefore increases its density. 

While the electrons are confined at the shock front by the mirror force of the shock-compressed field, they drift along the shock surface along the $-\hat{z}$ direction and are energized via SDA by the motional electric field $\vec{E}_0\parallel \hat{z}$. 
As a result of the SDA process, the reflected electrons increase their momentum along the direction of the upstream background  field, as revealed by the electron phase diagrams in Fig. \ref{fig:shockstruct}(b)-(d). 
In the \texttt{reference} run, the upstream background magnetic field is nearly along the $+\hat{y}$ axis.  
Correspondingly, we notice a number of electrons having large momenta along the $+\hat{y}$ direction ($p_y \gtrsim 3 \,m_e c$) far ahead of the shock front. 
Since the electrons accelerated by SDA gain momentum preferentially 
along the direction of the magnetic field, they induce an electron temperature anisotropy $T_{e\parallel}>T_{e\perp}$ in the upstream, over an extended region ahead of the shock (Fig. \ref{fig:shockstruct}(e)). 
Here $T_{e\parallel\ (\perp)}$ is the  electron temperature parallel (perpendicular) to the upstream magnetic field.

In quasi-perpendicular shocks, electrons are the only species that can propagate upstream  after being reflected by the shock front. 
The ions either advect downstream or are confined within a distance of $\sim r_{{\rm L},i}\simeq 80\, c/\omega_{pe}$ ahead of the shock. Beyond this distance, no shock-reflected ion is present, and the ion distribution is isotropic (as expected for the upstream medium at initialization).
This suggests that it is the
electron temperature anisotropy that triggers magnetic waves in the upstream, since the waves extend well beyond a few ion Larmor radii ahead of the shock (Fig. \ref{fig:shockstruct}(f)-(h)). 
Their wave vector is oblique with 
respect to the background upstream field 
(indicated by the white arrows in Fig. \ref{fig:shockstruct}(f)-(h)). The magnetic field fluctuations, $\delta\vec{B}\equiv  \vec{B}-\vec{B}_0$, grow
preferentially perpendicular to the plane defined by the wave vector and the background field, since 
$\delta B_z$ is stronger than $\delta B_x$ or $\delta B_y$. 
We find that the waves have phase velocity equal to the upstream flow velocity $\vec{u}_0 = -0.15c\ \hat{x}$ in the simulation frame, which implies that they are purely growing modes in the upstream comoving frame. 
In Section \ref{sec:waves}, we confirm that the waves are indeed generated by the electrons, 
and that they are due to the electron firehose instability.
The self-generated waves mediate the second stage of electron acceleration (beyond the initial SDA phase), in which the reflected electrons 
are scattered back towards the shock by the upstream waves and undergo multiple cycles of 
SDA, in a process similar to the Fermi mechanism. 
The energy gain of the accelerated electrons 
is dominated by multiple cycles of SDA, whereas the direct contribution from the interaction with the upstream waves is marginal.
 The trajectory and energy evolution of a typical electron undergoing Fermi-like acceleration
is shown in Fig. 8 of Paper I.

\section{Injection via Shock Drift Acceleration}\label{sec:SDA}
\begin{figure*}
\begin{center}
\includegraphics[width = 0.9\textwidth]{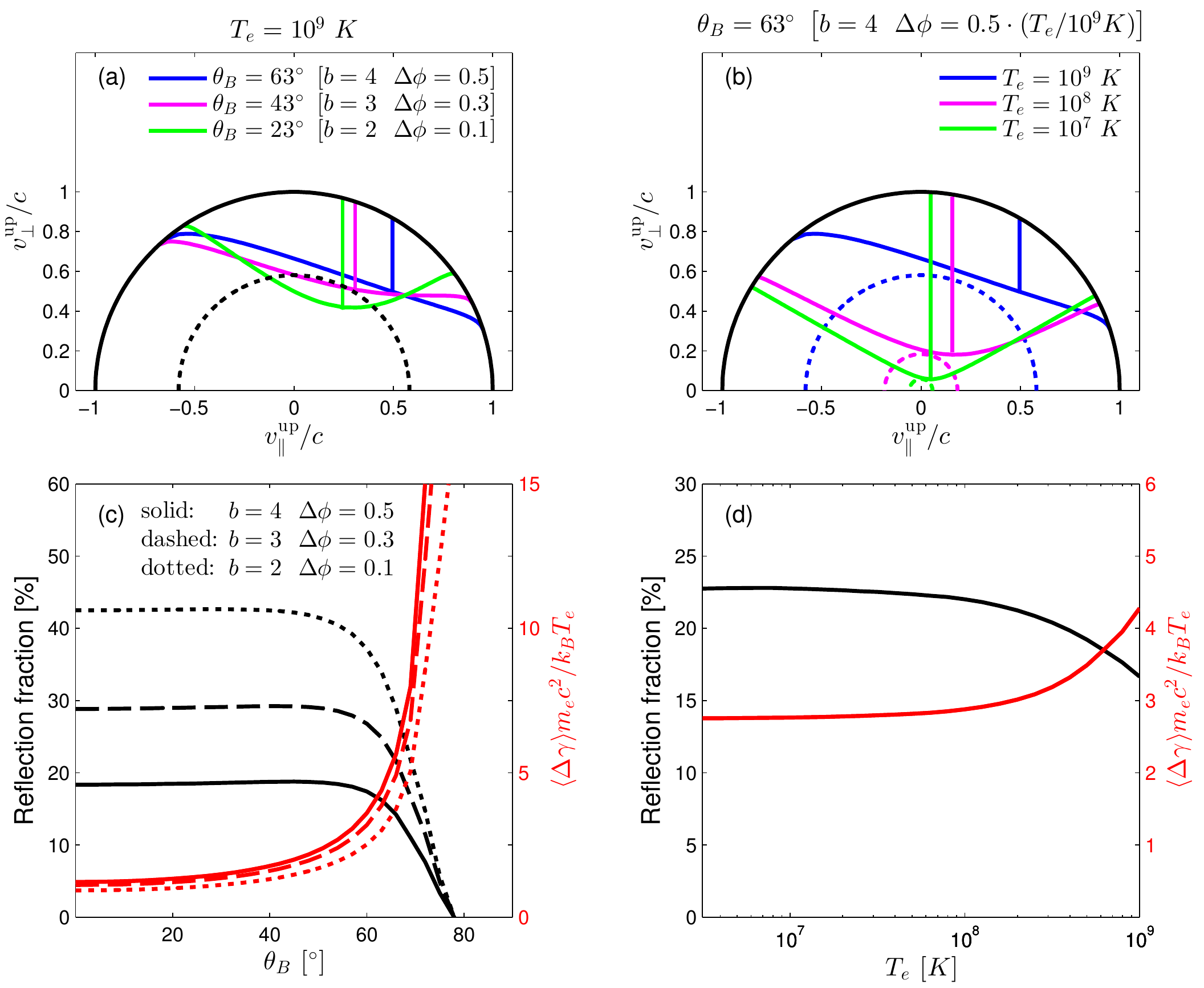}
\caption{Dependence of the SDA injection efficiency on the obliquity angle $\theta_B$ (on the left) and the electron temperature $T_e$ (on the right), for a low Mach number shock with $M_s = 3$. 
Panel (a) shows the region in velocity space ($v_{\parallel}^{\rm up}$-$v_{\perp}^{\rm up}$) where electrons populating a thermal 
distribution with $T_e = 10^9K$ are allowed to participate in SDA, as a function of the magnetic obliquity $\theta_B$.
The solid black semi-circle indicates the speed of light.
The vertical colored lines indicate the values of $u_t$ corresponding to different choices of $\theta_B$, as indicated in the legend. For a given $\theta_B$, the region allowed for 
SDA reflection is to the left of the colored vertical line and above the 
semi-horizontal line of the same color  (within the limit of the speed of light). 
The region to the right of the vertical colored line is the area in velocity space
that the SDA-reflected particles will occupy. 
The dashed black semi-circle indicates the electron thermal velocity $v_{{\rm th},e}\equiv \sqrt{2k_B T_e/m_e}$ for $T_e = 10^9K$. 
The overlap between the region near the thermal velocity semi-circle and the allowed region for reflection indicates the SDA efficiency: the more they overlap, the higher the number of electrons participating in SDA, thus the higher the SDA efficiency. In the examples shown here the efficiency increases with decreasing $\theta_B$. 
Panel (b) shows a similar diagram investigating the dependence 
of the SDA efficiency on the electron temperature $T_e$. We fix 
 $\theta_B = 63^\circ$ and $b=4$ and scale the cross-shock potential as $\Delta\phi = 0.5\, (T_e /10^9K)$. Here the dashed colored semi-circles indicate the electron thermal velocity at different temperatures (color coding in the legend).  For different $T_e$, the solid colored lines mark the boundaries of the region allowed for SDA reflection (to the left of the vertical line) and of the region occupied by the SDA-reflected electrons (to the right of the vertical line).
Panel (c) shows the reflection fraction (left axis, in black) and average energy gain (right axis, in red) as a function of $\theta_B$, for three representative choices of $b$ and $\Delta\phi$. 
Panel (d) shows the reflection fraction (tick marks 
on the left axis) and average energy gain (tick marks on the right axis) as a function of $T_e$, with $\theta_B = 63^\circ, \, b = 4$ and $\Delta\phi = 0.5 \,(T_e/10^9 K)$. 
} 
\label{fig:SDAtheory}
\end{center}
\end{figure*}

Since SDA plays a major role in the electron Fermi-like acceleration process summarized above, it is important to understand the main properties of SDA: (\textit{i}) its efficiency, i.e., the fraction of 
electrons from a thermal distribution that will be reflected upstream by SDA (and thereby injected into the acceleration process); and (\textit{ii}) the energy gain from each cycle of SDA. Our fully relativistic theory of SDA has been presented in detail in Section 4.2.1 of Paper I. 
In this section, we briefly summarize our previous findings and focus on 
the dependence of the SDA injection efficiency on the magnetic field obliquity angle $\theta_B$
and the electron temperature $T_e$. 

SDA is only viable in subluminal shocks. In contrast, at superluminal shocks, the velocity required to boost from the upstream rest frame to the de-Hoffman-Teller (HT) frame \citep{HT1950}
\begin{equation}\label{eq:ut}
u_t =  u_{\rm sh}^{\rm up}\sec\theta_B = \sqrt{\frac{2\Gamma k_BT_e}{m_e } }\sqrt{\frac{m_e}{m_i}}M_s\sec\theta_B\  
\end{equation}
exceeds the speed of light. In superluminal shocks, no particle travelling along the magnetic field towards the upstream can outrun the shock, so the injection efficiency is expected to vanish. Therefore, in this work, we focus only on electron acceleration in subluminal shocks. We confirm in Section \ref{sec:angles} that in superluminal shocks the SDA process cannot operate, and the electron acceleration efficiency vanishes. 

The efficiency of SDA decreases with increasing $u_t$, as the 
minimal electron energy required to participate in the SDA process increases with $u_t$ (see Section 4.2.1 in Paper I). 
In addition, the efficiency of SDA depends on the cross-shock electrostatic potential 
 and the magnetic field compression at the shock.
Inheriting the notation used in Paper I, we use $\Delta\phi$
to denote the change of potential energy of an electron as it crosses the shock
from upstream to downstream, in units of its rest mass energy. We indicate with   $b$ the compression of the magnetic field strength $B/B_0$ at the shock front.\footnote{We remark that $b$ is typically larger than the value predicted by the Rankine-Hugoniot relations. 
As seen from the shock structure of the \texttt{reference} run, for quasi-perpendicular shocks the magnetic field compression is higher than in the far downstream (both the density and the transverse magnetic field show an overshoot at the shock front), thus $b$ is larger than predicted by the Rankine-Hugoniot relations. 
On the other hand, in quasi-parallel shocks $b$ approaches the Rankine-Hugoniot relation for the compression of the magnetic field strength, as the overshoot at the shock front is less prominent.}
A decrease in $b$ or an increase in $\Delta\phi$ 
have the effect of increasing the minimum pitch angle required for SDA reflection in the HT frame. This leads to a lower fraction of incoming electrons that can participate in the SDA process.

The fractional energy gain from a single cycle of SDA increases monotonically with $u_t$ and is independent of $\Delta\phi$ or $b$ (see Equation (24) of Paper I)
\begin{equation}
\frac{\Delta\gamma}{\gamma_{i}^{\rm up}} \equiv  \frac{\gamma_{r}^{\rm up}-\gamma_{i}^{\rm up}}{\gamma_{i}^{\rm up}} =\frac{2u_t\left(u_t - v_{i\parallel}^{\rm up}\right)}{c^2 - u_t^2}\ .\label{eq:egain}
\end{equation}
Here $\gamma_{i(r)}^{\rm up}$ is the electron Lorentz factor before (after, respectively) the SDA cycle, and $v_{i\parallel}^{\rm up}$ is the particle velocity parallel to the magnetic field before SDA. All the quantities are measured in the upstream rest frame.

To understand the dependence of the SDA process on the upstream conditions, it is useful to illustrate how the key parameters of SDA, namely $u_t$, $\Delta\phi$ and $b$, 
depend on the pre-shock properties. 
Equation \eqref{eq:ut} shows that, for fixed $m_i/m_e$ and $M_s$, the HT velocity $u_t$ increases with $T_e$ and $\theta_B$. 
In addition, the cross shock potential $\Delta\phi$ scales roughly as \citep{Amano2007}
\begin{equation}\label{eq:deltaphi}
\Delta \phi  \sim   \frac{m_i u_0^2}{2\,m_e c^2}=M^2 \Gamma \frac{k_B T_e}{m_e c^2}\ .
\end{equation}
The value of $\Delta\phi$ also depends on $\theta_B$ for fixed $M$ and $T_e$, 
which is not accounted for by the equation above.
Intuitively, this can be understood as follows. The cross-shock potential develops
as a result of the excess of ions in the overshoot formed at the shock front \citep[see e.g.][]{Gedalin2004}.
At smaller obliquities, the magnetic barrier of the shock-compressed field is weaker, so the upstream
ions can  advect downstream more easily along the magnetic field. It follows that the magnetic overshoot at the shock front is smaller in quasi-parallel shocks (i.e., $b$
 decreases with decreasing $\theta_B$). Since  the cross-shock potential is related to the overshoot in ion density, $\Delta\phi$ is expected to decrease at lower $\theta_B$, for fixed $M$ and $T_e$.
An analytical theory of the exact dependence of $\Delta\phi$ and $b$ on the magnetic obliquity is beyond the scope 
of this paper. In the upcoming sections, we shall use the 
values of $\Delta\phi$ and $b$ measured from our simulations. 

To illustrate the effect of various pre-shock conditions, we show in Fig. \ref{fig:SDAtheory} how the SDA reflection fraction (i.e., the injection efficiency) and the average energy gain vary as a function of $\theta_B$ and $T_e$. The results in Fig. \ref{fig:SDAtheory}  are 
based on the analytical model of SDA presented in Paper I. 
Fig. \ref{fig:SDAtheory}(a) and (b) identify the region allowed for SDA reflection in the velocity space $v_{\parallel}^{\rm up}- v_{\perp}^{\rm up}$, where 
$v_{\parallel(\perp)}^{\rm up}$ is the 
velocity component parallel (perpendicular, respectively) to the ambient upstream magnetic field. 
The solid black circle indicates the limit of the speed of light. 
The electron thermal velocity $v_{{\rm th},e} \equiv \sqrt{2k_B T_e/m_e}$ is indicated by the dashed semi-circles.
The colored vertical lines denote $v_{\parallel}^{\rm up} = u_t$, where $u_t$ is determined by Equation \eqref{eq:ut}. 
To be reflected at the shock via SDA, an incoming electron has to move 
towards the shock, i.e., $v_{\parallel}^{\rm up}<u_t$, and its transverse velocity $v_{\perp}^{\rm up}$ should be larger than a critical value that depends on $v_{\parallel}^{\rm up}$, $b$ and $\Delta\phi$. This critical threshold is indicated by the colored solid curve to the left of the vertical line having the same color. The area bounded by these two limits, together with the speed of light (i.e., within the black solid circle), indicates the region in velocity space allowed for SDA reflection.
The overlap between the region near the thermal velocity semi-circle (dashed line) and the allowed region for reflection provides an estimate of the SDA efficiency: the more they overlap, the larger the number of electrons participating in SDA, so the higher the SDA efficiency.
The region bounded by the colored curve to the right of the colored vertical line  indicates the 
area in velocity space that the electrons occupy after SDA reflection. 

Fig. \ref{fig:SDAtheory}(a) shows the effect of varying 
$\theta_B$ at fixed Mach number ($M_s=3$) and electron temperature ($T_e = 10^9K$). We use the values of $b$ and $\Delta\phi$ measured from our simulations, for different choices of $\theta_B$. 
At higher $\theta_B$, the HT velocity $u_t$ increases (see Equation \eqref{eq:ut}), and in fact the vertical colored lines in Fig. \ref{fig:SDAtheory}(a)  --- corresponding to $v_{\parallel}^{\rm up} = u_t$ --- shift to the right. With larger $\theta_B$, we find from our simulations that both $\Delta\phi$ and $b$ increase (see the legend in Fig. \ref{fig:SDAtheory}(a)).
The increase in $\Delta\phi$ raises the minimum energy required for reflection and thus decreases the SDA efficiency. 
On the other hand, the increase in $b$ allows particles with a wider range of pitch angles to participate in SDA (see Section 4.2.1 in Paper I), thus balancing the effect of $\Delta\phi$ to some extent. 

The combined effect of $\Delta\phi$ and $b$ is illustrated in Fig. \ref{fig:SDAtheory}(c), where we show the SDA injection efficiency as a function of $\theta_B$.
The three black curves (dotted, dashed and solid, with tick marks on the left axis) illustrate the dependence of the injection efficiency on $\theta_B$, for three representative combinations of $b$ and $\Delta\phi$, as indicated in the legend. In our simulations, we find that at lower obliquities, the values of both $\Delta\phi$ and $b$ tend to decrease, so one should shift from the solid to the dashed and then  to the dotted curve, for lower $\theta_B$.
We find that the injection efficiency drops when $\theta_B\gtrsim 60^\circ$, i.e., when the shock becomes quasi-perpendicular, and it vanishes near $\theta_B \simeq 78^\circ$. This is because the shock becomes superluminal, i.e., $u_t>c$ in Equation \eqref{eq:ut}, for our choice of $m_i/m_e = 100$, $M_s=2$ and $T_e=10^9 \, K$.  

As discussed earlier, superluminal shocks are poor particle accelerators, since particles streaming along the field toward the upstream cannot outrun the shock, and the Fermi process is suppressed. However, in the context of the SDA theory, we find that electron acceleration becomes inefficient already for $u_t \gtrsim v_{{\rm th},e}$, which is more stringent than the superluminality criterion $u_t > c$. When $u_t \gtrsim v_{{\rm th},e}$, most 
of the thermal electrons cannot participate in the SDA process, so the injection efficiency becomes negligible.  
For the case of hot electrons ($T_e=10^9\, K$) considered in Fig. \ref{fig:SDAtheory}(a) and (c), 
the two criteria are similar, since the thermal velocity $v_{{\rm th},e}\sim 0.6\,c$ is quasi-relativistic. 
Yet, when studying colder electrons whose thermal velocity is non-relativistic, it is important to consider that the electron acceleration efficiency is already expected to decrease when $u_t \gtrsim v_{{\rm th},e}$.

The red curves in Fig. \ref{fig:SDAtheory}(c) illustrate the dependence of the average electron energy gain $\langle\Delta\gamma\rangle m_e c^2/k_B T_e$ on the field obliquity (tick marks on the right axis). The mean energy gain increases monotonically with $\theta_B$, because the SDA fractional energy gain increases with $u_t$ (see Equation \ref{eq:egain}), and $u_t$ grows monotonically with $\theta_B$ (see Equation \ref{eq:ut}).  
We point out that the average is measured over a population of electrons 
following a thermal distribution with $T_e = 10^9K$ (as expected for the upstream electrons at initialization), which is appropriate for the first SDA cycle. 
When the electrons undergo further SDA cycles, the exact value of the mean energy gain will change, since the
reflected electrons no longer follow a thermal distribution. However, the trend of higher  energy gains for larger $\theta_B$
 should remain the same. 

Similarly, Fig. \ref{fig:SDAtheory}(b) and (d) illustrate the effect of varying $T_e$ on the SDA efficiency and the average energy gain at quasi-perpendicular ($\theta_B = 63^\circ$) low Mach number ($M_s = 3$) shocks. For 
the calculations presented in these two panels, we use a fixed value of the magnetic compression ratio ($b=4$)  and we scale the cross-shock potential with the electron temperature such that $\Delta\phi =0.5\,(T_e/10^9K)$, as suggested by Equation \eqref{eq:deltaphi} and verified in our simulations. In the low temperature regime ($T_e\lesssim 10^8 K$) most relevant for the ICM, the SDA efficiency 
and the electron energy gain (in units of $k_B T_e$) depend weakly on the electron temperature. The injection efficiency stays around $23\%$, while the  mean electron energy gain is $\langle\Delta\gamma\rangle m_e c^2/k_B T_e\sim 3$. At higher temperatures ($T_e\gtrsim 10^8 K$), 
the efficiency drops slowly with increasing $T_e$, while the electron energy gain (normalized to the thermal energy $k_B T_e$) increases with $T_e$.

In summary, at fixed electron temperature, the SDA
efficiency decreases at higher magnetic obliquities, $\theta_B$, 
while the average energy gain increases with $\theta_B$.
At fixed $\theta_B$, both the SDA efficiency and the 
average energy gain depend weakly on $T_e$ in the low temperature regime ($T_e\lesssim 10^8 \, K$) most relevant for galaxy clusters. 
When the temperature rises beyond $T_e \gtrsim 10^8 K$, 
the SDA efficiency decreases and the average energy gain moderately increases at higher electron temperatures.

\vspace{0.3in}

\section{The Upstream Waves}
\label{sec:waves}
\begin{figure}
\begin{center}
\includegraphics[width = 0.7\textwidth]{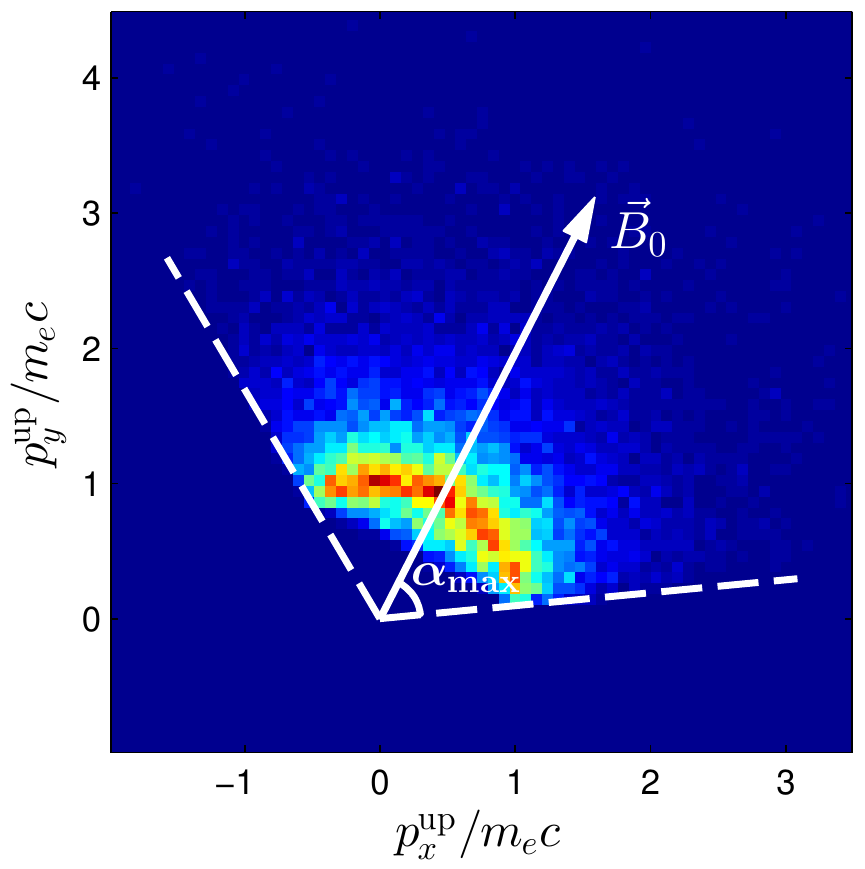}
\caption{Momentum space $p_x$-$p_y$ of the electron beam in our model, with $\theta_B = 63^\circ, \,\alpha_{\rm max}=57^\circ,\, \gamma_{\rm min}-1=0.4,\, \gamma_{\rm max}-1=6.7$. The momentum of the beam is centered around the direction of the background magnetic field, indicated by the white solid arrow. The beam electrons are distributed uniformly in solid angle, with pitch angle ranging from $0$ to $\alpha_{\rm max}$, as bounded by the dashed lines.}
\label{fig:beamsetup}
\end{center}
\end{figure}

In this section, we investigate in detail the 
properties of the upstream waves that mediate
the Fermi-like acceleration process.
We first confirm that the waves 
are triggered by the electrons returning upstream after the SDA process and show that the waves are associated with the oblique mode of the electron firehose instability. 
We then investigate how the generation
of the upstream waves depends on the plasma conditions.

\subsection{Setup for Periodic-Box Simulations}
To explore the physics of the upstream waves, we have performed 2D PIC simulations on a square computational domain in the upstream rest frame, with periodic boundary conditions along the $x$ and $y$ directions (periodic box simulations, hereafter). The simulations are targeted to capture the evolution of the upstream medium far ahead of the shock, in the upstream fluid frame.   The plasma in the box consists of a background electron-ion plasma
and an electron beam. The background plasma follows the thermal distribution initialized in our shock simulations. The electron beam mimics the properties of the SDA-reflected electrons, based on the SDA injection model discussed in Section  \ref{sec:SDA}. In this computational setup, the free energy in the electron beam is the only available  source of instability.

According to the prediction of SDA, a certain 
fraction of the thermal electrons propagating toward the shock are reflected back upstream. 
As result of SDA, the reflected electrons have higher energy and smaller pitch angles. 
To mimic the properties of the SDA-reflected electrons, we extract four parameters from the SDA injection model described in Section  \ref{sec:SDA}: the fraction of electrons from the upstream thermal distribution that satisfy the reflection condition, $n_{\rm ref}/n_e$ (normalized to the number density of the background electrons); the maximum pitch angle of the reflected electrons, $\alpha_{\rm max}$; the minimum and maximum kinetic energy of the reflected electrons, $\gamma_{\rm b,min}-1$ and $\gamma_{\rm b,max}-1$. 

We set up our periodic box simulations as follows. 
For a given shock simulation listed in Table \ref{table:params}, 
the corresponding periodic box experiment
contains a background electron-ion plasma with the same temperature and magnetic field strength as in the shock simulation. Since there is no shock in the periodic box simulations, the orientation of the magnetic field is arbitrary (in our shock simulations, the magnetic field direction was defined with respect to the shock normal, aligned with $x$). Yet, we decide to orient the ambient magnetic field with respect to the $x$ axis at the same angle $\theta_B$  as in the shock simulations, 
for easier comparison.\footnote{Since all the shocks in Table \ref{table:params} are non-relativistic, the change 
of magnetic obliquity when transforming from the simulation frame to the upstream rest frame is negligible.} 

In addition to the background plasma, we initialize  an electron beam whose number density is a fraction $n_{\rm ref}/n_e$ of the electron density in the background. The  beam electrons follow a power-law distribution in kinetic energy in the range 
$\gamma_{\rm b,min}-1\leq \gamma_{\rm b}-1\leq \gamma_{\rm b,max}-1$, with a slope of $-4$.\footnote{We point out that the power-law distribution with a slope of $-4$ chosen for the beam electrons is just a convenient way to represent the energy spread of the SDA-reflected electrons. It should be distinguished from the power-law fit of the  electron energy spectra measured in the shock simulations, which gives a spectral index $p\simeq 2.4$ below the exponential cutoff (see Paper I). Also, we remark that for a slope as steep as $-4$, our results are insensitive to the exact value of the high-energy cutoff $\gamma_{\rm b,max}-1$.
}
The beam electrons are distributed isotropically in solid angle within a cone whose axis is aligned with the ambient magnetic field. The opening angle of the cone (or equivalently, the maximum electron pitch angle) is chosen to be $\alpha_{\rm max}$.\footnote{As shown in Fig. \ref{fig:SDAtheory}(a)-(b), particles having small pitch angles (or equivalently, $v_{\parallel}^{\rm up}/v_{\perp}^{\rm up}\ll1$) are outside the region occupied  by SDA-reflected electrons (delimited by the colored curves to the right of the colored vertical lines). Yet, since the solid angle is small near $\alpha = 0^\circ$, we still choose $[0^\circ, \alpha_{\rm max}]$ as a convenient proxy for the range of pitch angles of the reflected electrons.} To ensure charge and current compensation in our periodic simulations, we balance
the negative charge of the beam electrons with a corresponding excess of background ions. The beam current is neutralized by initializing the background electrons with a small bulk velocity. In our shock simulations, the charge and current imbalance introduced by the beam of returning electrons in the upstream is compensated self-consistently on short time scales (a few $\omega_{pe}^{-1}$).

For our reference run, using $b=4$ and $\Delta\phi = 0.5$ as input parameters to our SDA injection model, we obtain $n_{\rm ref}/n_e=0.18$, $\alpha_{\rm max} = 57^\circ$, $\gamma_{\rm b,min} - 1= 0.4$ and $\gamma_{\rm b,max} -1= 6.7$.  Fig. \ref{fig:beamsetup} shows the $p_x$-$p_y$ momentum space of the electron beam for our reference periodic box run.
We note that the beam electrons have a large momentum component along the direction parallel to the magnetic field (indicated by the white solid arrow).
When combined with the background isotropic electrons, this will introduce an electron temperature anisotropy $T_{e\parallel}>T_{e\perp}$ in the beam-plasma system, which is essential for triggering the upstream waves that we discuss below.

As regards to numerical parameters, we employ $5$ cells per electron skin depth on a square domain of $768\times 768$ cells. Since the upstream waves we observe in the shock simulations have small amplitudes $|B_z|/B_0\sim 0.1$ (Fig. \ref{fig:shockstruct}),  the noise level needs to be very low, in order to clearly resolve their exponential growth and measure the growth rate. To achieve such a high accuracy, we employ a large number of particles per cell ($512$ per species, for the background electrons and ions).

\begin{figure}[tbp]
\begin{center}
\includegraphics[width=0.6\textwidth]{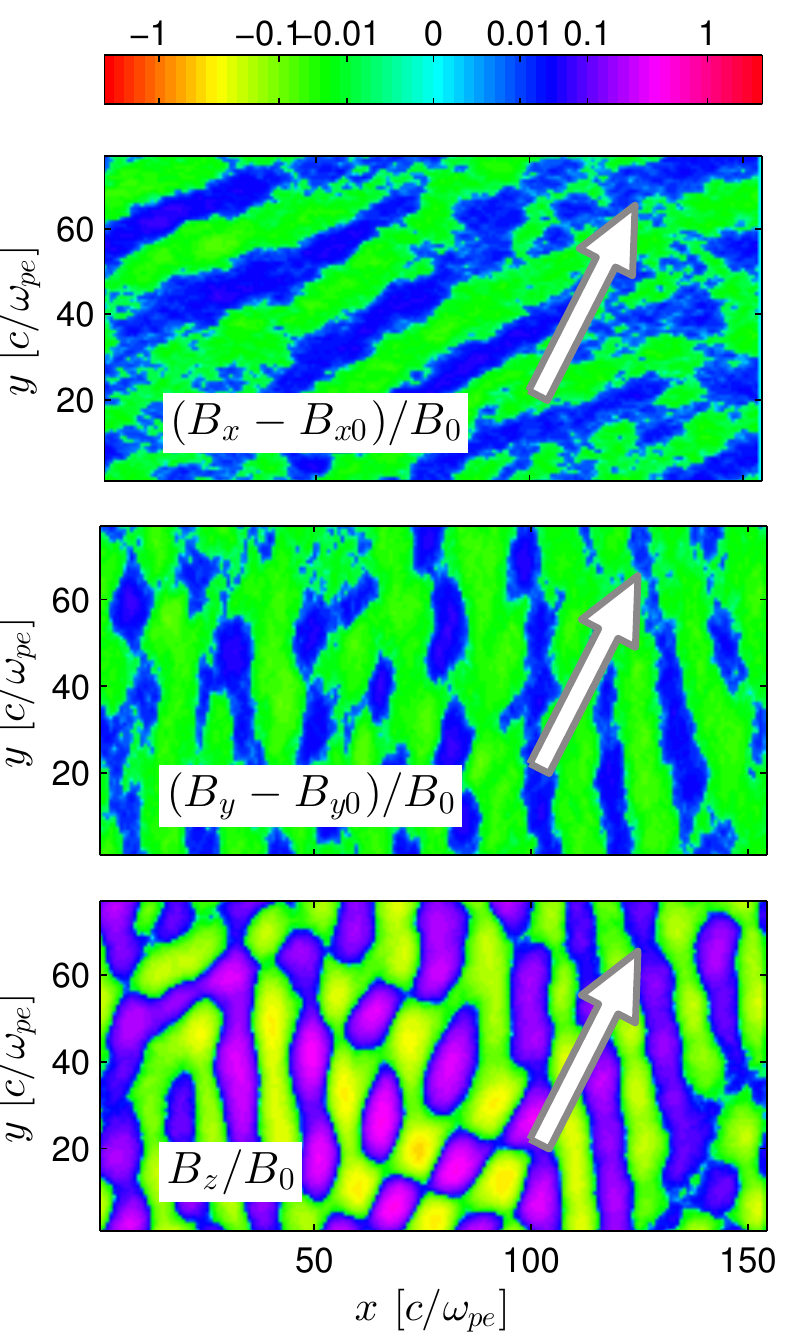}
\caption{ 2D plot of the magnetic waves in the reference \textit{periodic box simulation}. 
Note their similarity to the upstream waves in the reference \textit{shock simulation} (Fig. \ref{fig:shockstruct}(f)-(h)), in terms of wavelength and orientation of the wavevector. The white arrows indicate the orientation of the background magnetic field.
}
\label{fig:weibeltheta63}
\end{center}
\end{figure}

\subsection{Electron Oblique Firehose Instability} 
The setup of our periodic box simulations, with the electron 
temperature anisotropy $T_{e\parallel}>T_{e\perp}$ induced 
by the beam of SDA-reflected electrons, is unstable to the electron firehose instability. 
\cite{Hollweg1970} first discovered that high-beta plasmas with $T_{e\parallel}>T_{e\perp}$ are unstable to waves propagating along the background field (the so-called parallel firehose instability).
\cite{Paesold1999} demonstrated that the maximum growth rate of modes associated with electron anisotropies $T_{e\parallel}>T_{e\perp}$ in high-beta plasmas is attained at oblique  angles, i.e. $\vec{k}\times\vec{B}_0\neq 0$, where $\vec{k}$ is the wave vector.
Later, it was discovered that the oblique mode is purely growing (i.e., with zero real frequency), and its 
growth rate $\Gamma_{\rm max}$ lies in the range $\Omega_{ci}\ll\Gamma_{\rm max}\lesssim\Omega_{ce}$, where $\Omega_{ci}$ and $\Omega_{ce}$ are the ion and electron cyclotron frequencies, respectively \citep{Li2000,Gary2003,Camporeale2008}. Thus, the oblique mode grows faster than the parallel mode, whose growth rate is $\Gamma_{\rm max}\lesssim\Omega_{ci}$ \citep[e.g.][]{Davidson1984,Yoon1990,Yoon1995,Kunz2014}. Also, the threshold 
of the oblique mode is lower than that of the parallel mode. 
Due to its faster growth rate and lower threshold, the oblique mode of the electron firehose instability is
usually the dominant mode for anisotropic ($T_{e\parallel}>T_{e\perp}$) moderately magnetized plasmas, unless the wavevector is forced to align with the magnetic field \citep{Gary2003}.

Given the above expectations, there are two major aspects
we wish to investigate using our 2D periodic box simulation 
with parameters appropriate to the \texttt{reference} shock run in Table \ref{table:params}. 
First, we want to verify that our periodic box simulation can 
reproduce the waves we found in the 
shock simulation. 
Then, we want to check if the waves are indeed associated with the 
electron oblique firehose instability. 

The waves generated in our periodic box simulation 
are shown in Fig. \ref{fig:weibeltheta63}.
We find that their pattern closely resembles what we observed
in the corresponding shock simulation (Fig. \ref{fig:shockstruct}(f)-(h)).
In particular, in both the shock and the periodic box simulations, the magnetic fluctuations $\delta\vec{B}$
are stronger along the $\hat{z}$ direction than in $\hat{x}$ or $\hat{y}$. This is expected for the oblique firehose instability, where the largest contribution to $\delta\vec{B}$ is  predicted to be perpendicular to the plane formed by $\vec{B}_0$ and $\vec{k}$ \citep{Gary2003}.
In both the shock and the periodic box simulations, the waves show two dominant wave-vectors, symmetric with respect to the ambient field $\vec{B}_0$. Their wavelength is $\sim 10 - 20 \,c/\omega_{pe}$, much smaller than the ion gyration radius ($r_{{\rm L},i}\sim 80 \,c/\omega_{pe}$ from Equation \eqref{eq:rLi}), which confirms that the waves are governed by the electron physics. 
The waves in the periodic box simulation have zero real frequency (i.e., they are non-propagating, as expected for the oblique firehose instability), which agrees with the fact that the upstream waves in our shock simulation move together with the upstream flow, as discussed in Section \ref{sec:shockstructure}. 

Given the close similarity between the waves  in our periodic box and in the shock simulations, and the fact that the periodic box experiment excludes other sources of instability --- except for the beam of SDA-reflected electrons ---  we conclude that the upstream waves in the shock simulation are generated by the electrons returning upstream after the SDA process.

We now demonstrate that  the properties of the waves in our periodic box simulation are fully consistent with the expectations of the electron oblique firehose instability. At initialization, the beam-plasma system in our reference periodic box is 
above the threshold for the oblique mode of the electron firehose instability (compare the red curve  with the horizontal black dashed line in Fig. \ref{fig:wavegrowth}(b)), 
\begin{equation}\label{eq:threshvalue}
1-\frac{T_{e\perp}}{T_{e\parallel}} - \frac{1.27}{\beta_{e\parallel}^{0.95}} > 0
\end{equation}
where $\beta_{e\parallel}\equiv 8\pi n_e k_B T_{e\parallel}/B_0^2$ is the electron beta parallel to the background magnetic field, and we adopt the instability threshold
derived by \cite{Gary2003}.
The growth rate of the instability, as measured from the exponential phase ($10\lesssim \Omega_{ce}t\lesssim50$) in Fig. \ref{fig:wavegrowth}(c) (red curve), is $\sim 0.05\ \Omega_{ce}$. 
The predicted growth rate for a system with $\beta_{e\parallel} = 10$ and $T_{e\perp}/T_{e\parallel} = 0.7$, similar to our setup, is $\sim 0.08\ \Omega_{ce}$ \citep{Camporeale2008}, which compares favorably with our result. 
\cite{Camporeale2008} further predict that the wavelength   
of the fastest growing mode is $\sim 15\, c/\omega_{pe}$ and the wave vector is oriented at $\sim 70^\circ$ with respect to $\vec{B}_0$. 
This is not very different from the wavelength ($\sim29\  c/\omega_{pe}$) and the wave vector orientation (at an angle of $\sim57^\circ$ with respect to $\vec{B}_0$) measured in our periodic box simulation (wave pattern shown in Fig. \ref{fig:weibeltheta63}). The agreement is reasonable, considering that our system does not match the particle distribution in \cite{Camporeale2008} exactly.
In addition, by running simulations with $m_i/m_e=100$ (our standard choice), $m_i/m_e=400$ and the realistic case $m_i/m_e=1836$, we have verified that the growth rate and dominant wavelength of the instability do not depend  on the ion-to-electron mass ratio $m_i/m_e$, as long as $m_i/m_e\gg1$. This is indeed expected for the oblique firehose instability \citep{Gary2003,Camporeale2008}, whereas for the parallel firehose instability $\Gamma_{\rm max}\propto m_e/m_i$.\footnote{We have directly tested the scaling $\Gamma_{\rm max}\propto m_e/m_i$ expected for the parallel mode by performing a suite of 1D simulations (with different $m_i/m_e$) where the computational box is aligned with $\vec{B}_0$.} The fact that the oblique mode is insensitive to the mass ratio explains why our choice of a reduced mass ratio $m_i/m_e = 100$ is sufficient to capture the electron acceleration physics, as we have demonstrated in Appendix B of Paper I. 
Over time, the temperature anisotropy decreases (so, $T_{e\perp}/T_{e\parallel}$ increases,  as shown by the red line in Fig. \ref{fig:wavegrowth}(a)) and the wave energy $\delta B^2/B_0^2$ grows (red curve in Fig. \ref{fig:wavegrowth}(c)). This is consistent with the fact that the instability is triggered by the free energy of the temperature anisotropy and that the waves scatter the electrons towards isotropization \citep[e.g.][]{Hellinger2014}.

Apart from the oblique firehose mode, other instabilities can be triggered by an electron temperature anisotropy $T_{e\parallel}>T_{e\perp}$: the parallel electron firehose instability 
and the ordinary-mode instability \citep[akin to the Weibel instability,][]{Ibscher2012}, whose properties are summarized in Table 3 of \cite{Lazar2014}. We can confidently identify our instability as the oblique firehose mode, for the following reasons. 
The fact that the dominant mode of our instability is purely growing rules out the parallel firehose instability, which has a non-zero real frequency. Also, the growth rate of the parallel firehose mode should depend on the ion-to-electron mass ratio, which is not the case in our simulations. In addition, the growth rate of our instability $(\lesssim\Omega_{ce})$ is incompatible with the expectations from the ordinary-mode instability, whose growth rate is $>\Omega_{ce}$. The fact that we do not observe strong fluctuations in the electric field component parallel to $\vec{B}_0$ further argues against the ordinary-mode instability.\footnote{We point out that the whistler waves discussed in \cite{Riquelme2011} are triggered when $T_{e\parallel}<T_{e\perp}$, whereas the opposite temperature anisotropy (i.e., $T_{e\parallel}>T_{e\perp}$) is present in the upstream region of our shocks.}

In summary, the excellent agreement between our results and the theory of the oblique  firehose mode 
suggests that the upstream waves ahead of low Mach number shocks are triggered by the returning electrons  via the oblique firehose instability. We remark that, since the dominant mode is oblique with respect to both the shock normal and the upstream magnetic field, multi-dimensional shock simulations are of paramount importance to characterize the electron acceleration physics.

\subsection{Dependence of Wave Generation on Plasma Conditions}
\begin{table}
\begin{center}
\vspace{-0.1in}
\begin{tabular}{c c c c c }
\hline 
Run & $\frac{n_{\rm ref}}{n_e}$ & $\alpha_{\rm max}$ & $\frac{\gamma_{\rm b,min}-1}{k_B T_e}$ & 
$\frac{\gamma_{\rm b,max}-1}{k_BT_e}$\tabularnewline
\hline 
\hline 
\texttt{refb} & 0.18 & $57^{\circ}$ & 2.4 & 40\tabularnewline
\texttt{theta43b} & 0.40 & $72^{\circ}$ & 1.0 & 26\tabularnewline
\texttt{theta73b} & 0.02 & $31^{\circ}$ & 7.7 & 82\tabularnewline
\texttt{sig1e-1b} & 0.18 & $57^{\circ}$ & 2.4 & 40 \tabularnewline 
\texttt{sig1e-2b} & 0.18 & $57^{\circ}$ & 2.4 & 40 \tabularnewline
\texttt{Te1e8.0b} & 0.25 & $74^{\circ}$ & 1.8 & 31 \tabularnewline 
\hline 
\end{tabular}
\caption{Parameters Used for the Periodic Box Simulations}
\label{table:beam}
\end{center}
\end{table}

\begin{figure}[tbp]
\begin{center}
\includegraphics[width=0.8\textwidth]{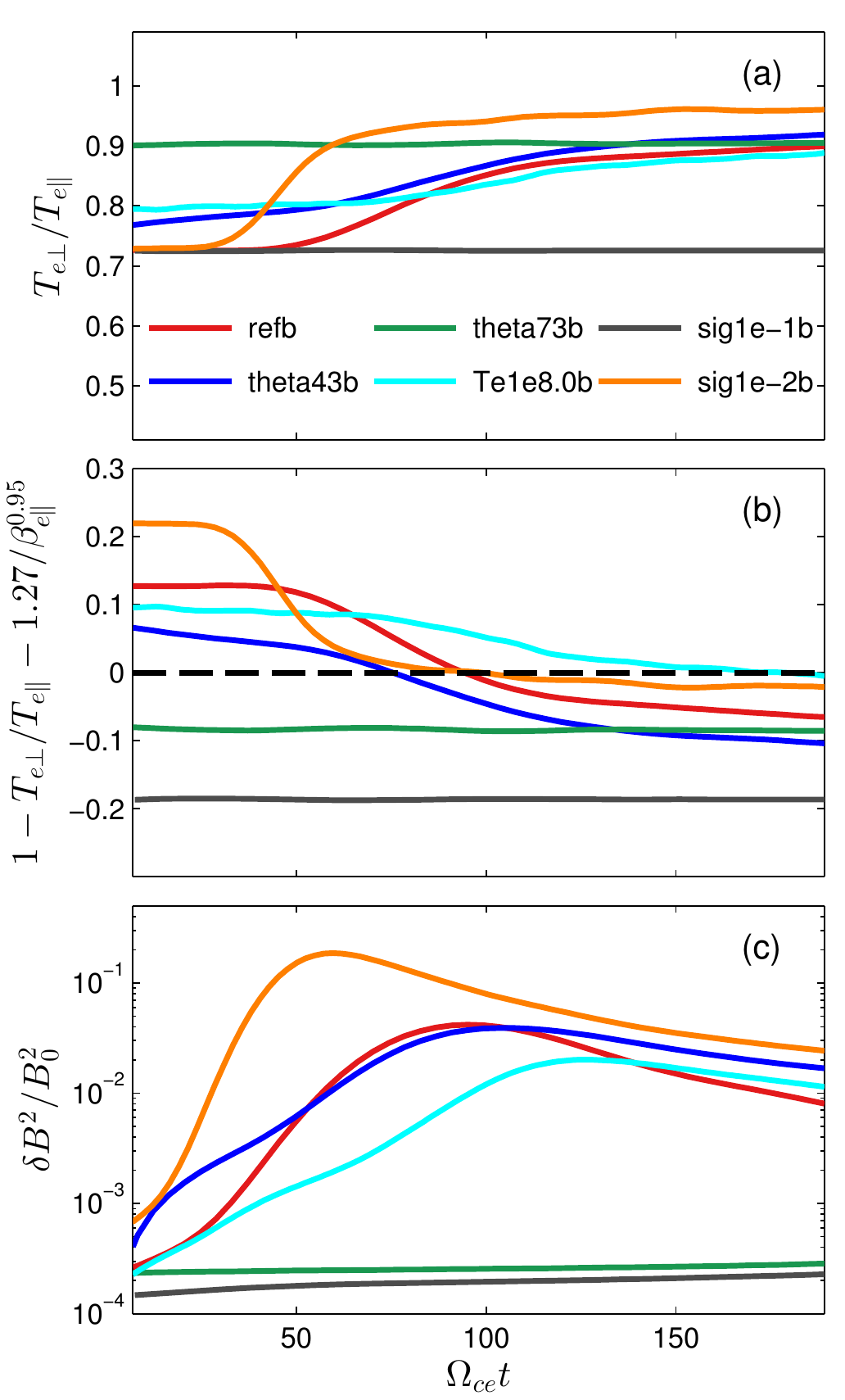}
\caption{Time evolution of various quantities measured in periodic box simulations with parameters listed in Table \ref{table:beam} and indicated in the legend of panel (a). Panel (a) traces the electron temperature ratio $T_{e\perp}/T_{e\parallel}$. 
Panel (b) follows the quantity $1-T_{e\perp}/T_{e\parallel}-1.27/\beta_{e\parallel}^{0.95}$ . The horizontal dashed line indicates the marginal stability threshold for a growth rate $\Gamma_{\rm max}=0.001\,\Omega_{ce}$, where $\Omega_{ce}$ is the electron cyclotron frequency.
Panel (c) traces the energy in the magnetic waves, normalized to the energy density of the background field. 
}
\label{fig:wavegrowth}
\end{center}
\end{figure}
To understand the conditions under which the electrons reflected by SDA can trigger the electron oblique firehose instability in the upstream region of low Mach number shocks, we perform five additional periodic box simulations. These are called \texttt{theta43b, theta73b, sig1e-1b, sig1e-2b, Te1e8.0b}, and they correspond to the shock simulations \texttt{theta43, theta73, sig1e-1\_63, sig1e-2\_63, Te1e8.0}. The beam parameters are listed in Table \ref{table:beam}.  
Note that the values of $n_{\rm ref}/n_e$, $\left(\gamma_{b,\rm min}-1\right)/k_BT_e$, $\left(\gamma_{b,\rm max}-1\right)/k_BT_e$ in Table \ref{table:beam}
reflect the trends discussed in Section \ref{sec:SDA}. In particular, the reflection fraction from SDA increases at lower $\theta_B$, while the energy gain of the reflected electrons decreases. The SDA reflection fraction increases slightly at lower temperatures, while the energy gain normalized to $k_B T_e$ moderately decreases for smaller $T_e$ (see Fig. \ref{fig:SDAtheory}). 

For the six periodic box simulations listed in Table \ref{table:beam}, Fig. \ref{fig:wavegrowth} shows the evolution of the electron temperature anisotropy $T_{e\perp}/T_{e\parallel}$ (panel (a)), the quantity $1-T_{e\perp}/T_{e\parallel}-1.27/\beta_{e\parallel}^{0.95}$ (panel (b)) that characterizes the departure from the instability threshold, and the wave energy $\delta B^2/B_0^2$ (panel (c)).

As apparent in Fig. \ref{fig:wavegrowth}, there is a clear dichotomy, depending on whether the electron anisotropy starts above or below the instability threshold (indicated with the horizontal black dashed line in panel (b)). 
For the two runs that start below the instability threshold, \texttt{theta73b} and \texttt{sig1e-1b},
the instability never grows (the green and gray curves in 
Fig. \ref{fig:wavegrowth}(c) stay at the noise level) and the temperature anisotropy remains constant (Fig. \ref{fig:wavegrowth}(a)).
Runs \texttt{theta73b} and \texttt{sig1e-1b} fail to exceed the instability 
threshold for different reasons.
In run \texttt{theta73b}, the temperature anisotropy is very weak ($T_{e\perp}/T_{e\parallel}\sim 1$), because the fraction of SDA-reflected electrons is very small. Even though SDA results in a substantial energy gain (third row in Table \ref{table:beam}), the reflected electrons have a negligible effect on the overall plasma anisotropy.
In contrast, for run \texttt{sig1e-1b}
the temperature anisotropy starts at the same level as in the reference run \texttt{refb}, where the 
instability does develop, as discussed in the previous subsection.
However, due to the strong magnetization ($\sigma = 0.1$, giving $\beta_p = 6$), the electron parallel beta $\beta_{e\parallel}\sim 3$ is too small  and the instability does not grow. 

For the simulations meeting the instability threshold at initialization, viz.,
\texttt{theta43b, Te1e8.0b, sig1e-2b}, 
the waves do grow (blue, cyan and orange curves in Fig. \ref{fig:wavegrowth}(c)) and they  isotropize the electron distribution (Fig. \ref{fig:wavegrowth}(a)), 
in analogy to the reference periodic box run \texttt{refb}. 
The saturation time of the wave energy (i.e., the end of the phase of exponential growth) is  roughly 
coincident with the time when the electron anisotropy falls below the instability threshold. These three runs meet the instability threshold, for the reasons that we now explain.
In run \texttt{theta43b}, although the beam electrons are less energetic
and populate a wider cone than  in the run \texttt{refb} ($\alpha_{\rm max} = 72^\circ$, as compared
to $\alpha_{\rm max} = 57^\circ$ in \texttt{refb}), the relative density 
of the electron beam is larger ($n_{\rm ref}/n_e = 0.4$, as compared to $0.18$ in \texttt{refb}). 
The combined effect is still to induce a strong temperature anisotropy $T_{e\perp}/T_{e\parallel}\simeq 0.75$,
similar to the \texttt{refb} run.
With comparable $\beta_{e\parallel}$ as in the \texttt{refb} run, the threshold is met. 
A similar argument applies to the case \texttt{Te1e8.0b}. 
In the run \texttt{sig1e-2b}, the beam has the same properties as in run \texttt{refb}, and thus $T_{e\perp}/T_{e\parallel}$ starts exactly at the level. Since the  magnetization is lower ($\sigma = 0.01$, as compared to $0.03$ in \texttt{refb}), 
the electron parallel beta is even higher than in the run \texttt{refb}, making 
it easier to exceed the threshold. 

In summary, the results of 
our periodic box simulations confirm the role of the threshold in Equation \eqref{eq:threshvalue} 
for the excitation of the electron oblique firehose instability.
In short, the pressure induced by the electron anisotropy 
should be stronger than the upstream magnetic pressure. 
This condition cannot be met when either the number density of 
returning electrons  is too low or the 
upstream plasma is too strongly magnetized. 

Finally, we point out that the upstream environment simulated in our periodic box runs differs in some respect from that in the shock simulations. In the shock simulations, a steady flow of electrons reflected via SDA is produced at the shock and propagates into the upstream, constantly driving the instability. 
In contrast, in the periodic box simulations, the electron anisotropy is set up at the initial time, and then the system relaxes towards isotropy as the instability grows. 
To investigate from first principles the dependence of the electron acceleration process on the pre-shock conditions, we still need 
to rely on fully-consistent shock simulations, as we describe in the next section.

\section{Dependence on the Pre-Shock Conditions}\label{sec:dependence}
In this section, we explore the dependence of electron acceleration 
in low-Mach number shocks on pre-shock conditions. To do this, we vary the magnetic obliquity angle $\theta_B$, the 
magnetization $\sigma$ and the electron temperature $T_e$, as listed in Table \ref{table:params}.
For completeness, we also comment briefly on the effect of varying the Mach number $M_s$, based on 
simulations presented in an earlier work \citep{Narayan2012}. 
All the electron energy spectra presented in this section are measured at the same time in units of $\Omega_{ci}^{-1}$ between  $60\,c/\omega_{pe}$ and $160\,c/\omega_{pe}$ ahead of the shock.

\subsection{Dependence on the Field Obliquity Angle $\theta_B$}\label{sec:angles}
\begin{figure*}[tbp]
\begin{center}
\includegraphics[height = 0.26\textheight]{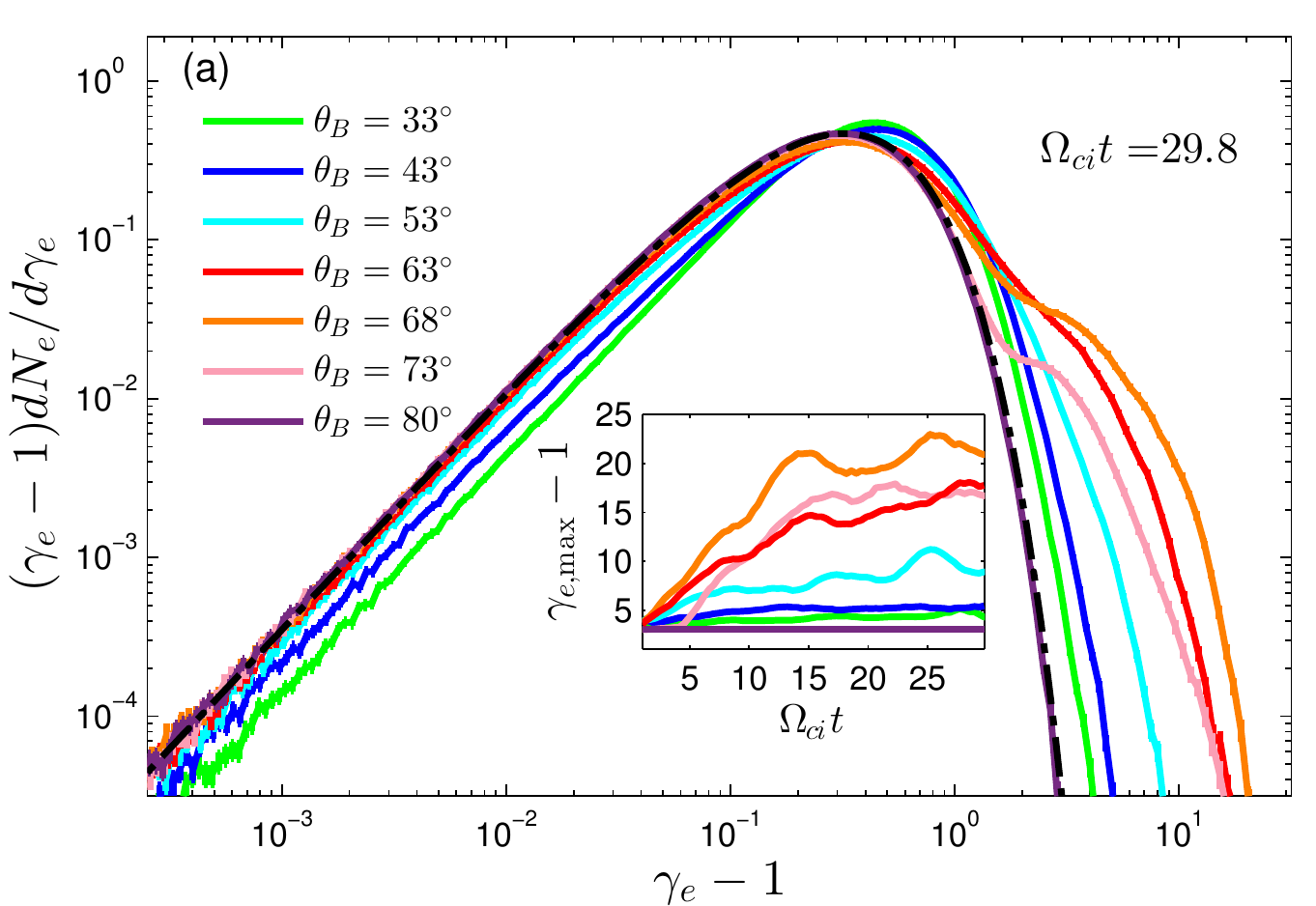}\includegraphics[height = 0.26\textheight]{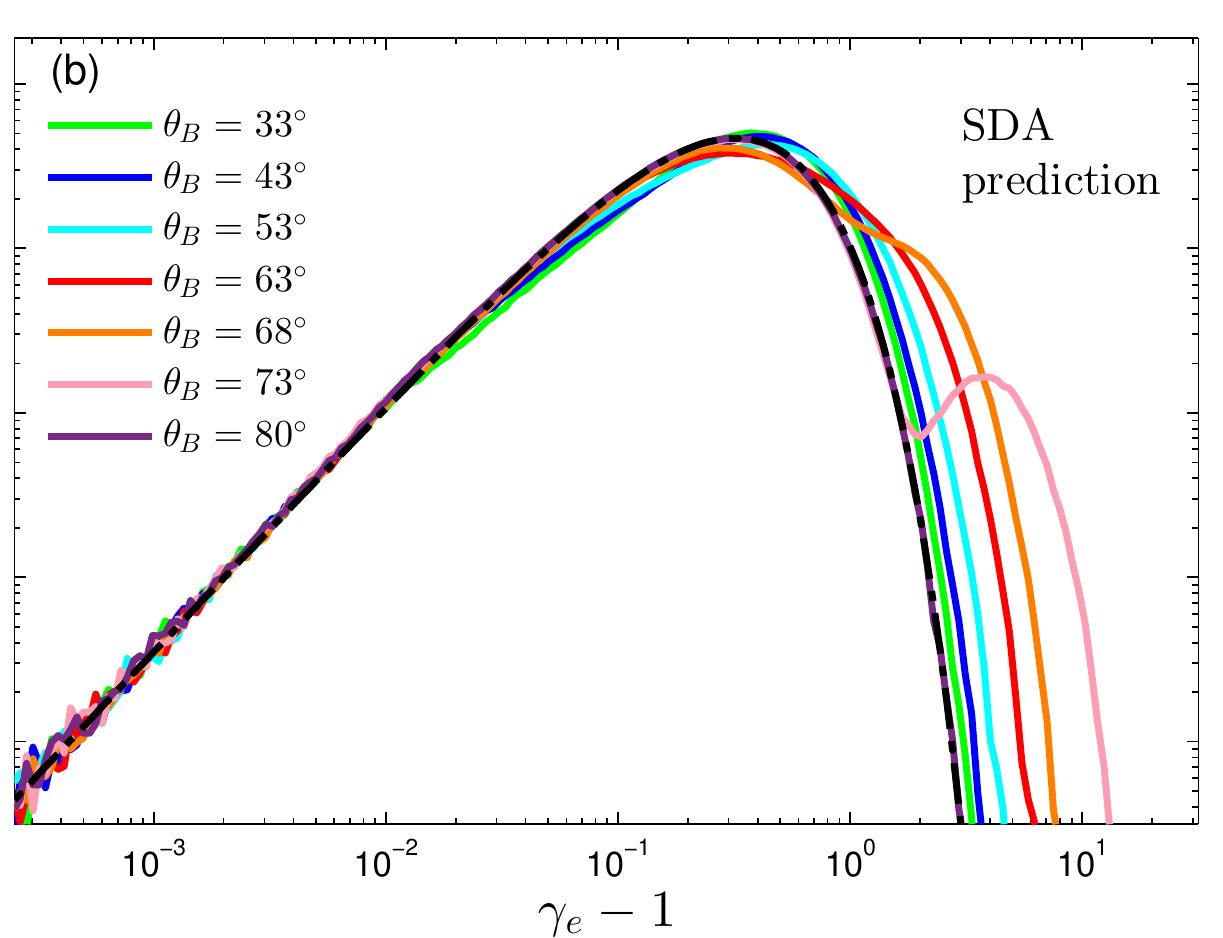}
\caption{Panel (a): upstream electron energy spectra measured at $\Omega_{ci}t=29.8$ in the runs \texttt{theta33, theta43, theta53, reference, theta68, theta73} and \texttt{theta80}, as indicated in the legend. The subplot shows the temporal evolution of the maximum energy of the upstream electrons. 
Panel (b): upstream electron energy spectra predicted by our SDA theory. 
In both panels, the black dot-dashed line corresponds to the drifting 
Maxwellian distribution at initialization, having $u_0 = 0.15\,c$ and 
$T_e = 10^9K$.
}
\label{fig:thetacompare}
\end{center}
\end{figure*}

We study the effect of the obliquity angle of the upstream
magnetic field, $\theta_B$, by comparing the
results from simulations with the same $T_e=10^9, M_s=3, \sigma=0.03$  but 
different $\theta_B$, in the range from $\theta_B = 13^\circ$ to $80^\circ$. In this section, we focus on runs 
having $\sigma=0.03$.
In Section \ref{sec:sigma}, we will briefly comment on the dependence on magnetic obliquity in shocks with $\sigma = 0.1$ and $\sigma=0.01$. 

The electron energy spectra measured at $\Omega_{ci}t = 29.8$ from the simulations  
\texttt{theta33, theta43, theta53, theta68, theta73, theta80}, 
along with the \texttt{reference} run (having $\theta_B = 63^\circ$),
are presented in Fig. \ref{fig:thetacompare}(a). 
As in Paper I, we define the maximum electron energy as
the Lorentz factor $\gamma_{e,\rm max}$ at which the particle 
number density drops below $10^{-4.5}$, the lowest level shown in our spectra.
The subpanel of Fig. \ref{fig:thetacompare}(a) traces the temporal evolution of the maximum energy of the upstream electrons. 
In  Fig. \ref{fig:thetacompare}(b), we show synthetic electron energy spectra obtained by adding a component of SDA-reflected electrons to the thermal electrons initialized in the upstream (see Section 4.2.2 of Paper I). The spectra in Fig. \ref{fig:thetacompare}(b) are based on our theory of SDA, considering a single acceleration cycle (hereafter, we shall call them SDA-synthetic spectra). So, they cannot account for the sustained Fermi-like acceleration process that is mediated by the oblique firehose waves. By comparing the left and right panels in Fig. \ref{fig:thetacompare}, we can quantify the importance of the firehose-mediated Fermi-like acceleration.

As discussed in Section \ref{sec:SDA}, no electron is expected to be accelerated via SDA in superluminal shocks, i.e., shocks with $u_t > c$ (Equation \eqref{eq:ut}). For $T_e = 10^9K, m_i/m_e = 100, M_s =3$, shocks with $\theta_B \gtrsim 78^\circ$ are superluminal. For this reason, the SDA-synthetic spectrum for $\theta_B = 80^\circ$
(purple curve in Fig. \ref{fig:thetacompare}(b)) overlaps with the electron energy spectrum at initialization (dot-dashed curve in Fig. \ref{fig:thetacompare}(b)), with no evidence for non-thermal electrons.
The energy spectrum measured from our simulation \texttt{theta80} also shows no sign of accelerated electrons (purple curve in Fig. \ref{fig:thetacompare}(a)). This indicates that, without injection via SDA, no electrons can be accelerated.  

Below the superluminal limit, the efficiency of SDA injection is expected
to increase with decreasing $\theta_B$, see Fig. \ref{fig:SDAtheory}(c). As discussed in Section \ref{sec:SDA}, the injection efficiency is still small if $u_t\gtrsim v_{{\rm th},e}$, since few of the thermal electrons will be fast enough to propagate back upstream, thus participating in the Fermi process. For $T_e = 10^9K, m_i/m_e = 100, M_s =3$, this limit correspond to an angle of $\theta_B\simeq 67^\circ$.
 Indeed, the electron non-thermal tails in shocks with $\theta_B$ below or close to this critical threshold contain a moderate fraction of electrons  ($10\%-20\%$), whereas the normalization of the non-thermal tail in the run \texttt{theta73} is very low ($\sim 2\%$).\footnote{The normalization of the non-thermal tail, a proxy for the electron injection efficiency, can be estimated from the point where the energy spectrum starts to deviate from a thermal distribution.}

\begin{figure*}[tbp]
\begin{center}
\includegraphics[width=1.0\textwidth]{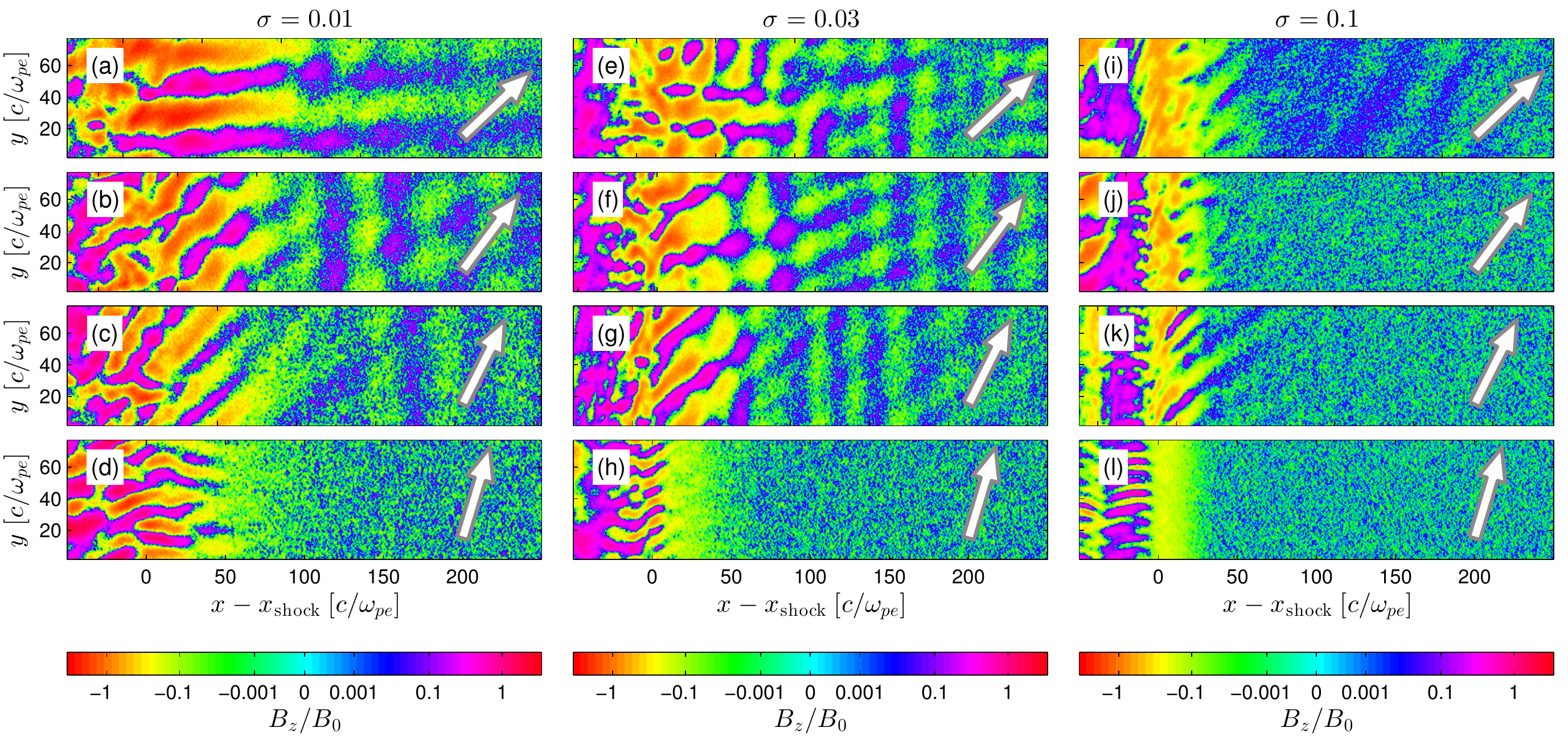}
\caption{
2D plots of $B_z/B_0$ at $\omega_{pe}t = 4275$ from runs 
with different magnetizations $\sigma$ and magnetic obliquities $\theta_B$. 
Panels (a) - (d): from runs with $\sigma = 0.01$ and different obliquities:
\texttt{sig1e-2\_43, sig1e-2\_53, sig1e-2\_63, sig1e-2\_73} from top to bottom;
Panels (e) - (h): from runs with $\sigma = 0.03$ and different obliquities: 
\texttt{theta43, theta53, reference, theta73} from top to bottom;
Panels (i) - (l): from runs with $\sigma = 0.1$ and different obliquities:
\texttt{sig1e-1\_43, sig1e-1\_53, sig1e-1\_63, sig1e-1\_73} from top to bottom.
In all the panels, the background field orientation in the upstream region is indicated with the white arrows. 
}
\label{fig:Bzsigs}
\end{center}
\end{figure*}

The average energy gain resulting from one cycle of SDA increases with $\theta_B$, as shown in Fig. \ref{fig:SDAtheory}(c) and suggested by the trend in the high-energy cutoffs of the SDA-synthetic spectra in Fig. \ref{fig:thetacompare}(b). However, by comparing panel (a) and (b), we find that the maximum energy $\gamma_{e,\rm max}$ in our shock simulations  \texttt{theta53, reference, theta68} has evolved to a value that is much larger than expected from a single cycle of SDA (compare the high-energy cutoffs between the two panels). This suggests that the Fermi-like acceleration mechanism is operating in those runs, and this explains the steady growth in the maximum electron energy shown in the subpanel of  Fig. \ref{fig:thetacompare}(a). In such shocks, the fraction of SDA-reflected electrons is large enough that the resulting electron temperature anisotropy in the upstream region can trigger the oblique firehose waves that mediate the Fermi-like process at late times (e.g., see the waves in runs \texttt{theta53, reference} shown in Fig. \ref{fig:Bzsigs}(f)-(g)). 

In contrast, the maximum energy of the electron spectrum from run \texttt{theta73} saturates soon after $\Omega_{ci}t\sim 15$ to a value almost identical to that of the SDA-synthetic spectrum ($\gamma_{e,\rm max}\sim 18$), indicating that the Fermi-like  acceleration mechanism does not operate in this case. This is consistent with the periodic box simulations in Section \ref{sec:waves}, where we have shown that with only $\sim 2\%$ of the incoming electrons being reflected via SDA  --- a value appropriate for the run \texttt{theta73} --- the  electron temperature anisotropy induced in the upstream   is too weak to trigger the electron firehose instability. 
In Fig. \ref{fig:Bzsigs}(h), we explicitly show that no upstream waves are present in the upstream region for run \texttt{theta73}. In the absence of upstream waves, the electrons cannot undergo multiple cycles of SDA, and the maximum energy saturates to the value predicted by one cycle of SDA.

In the quasi-parallel regime ($\theta_B\lesssim 45^\circ$), the reflection fraction is expected to be relatively high, while the average energy gain from each SDA cycle is only a few times the electron thermal energy (Fig. \ref{fig:SDAtheory}(c)). 
Thus, we expect that after one cycle of SDA, the electron energy spectra will not differ significantly from the drifting Maxwellian at initialization (see the green and blue curves in Fig. \ref{fig:thetacompare}(b), for SDA-synthetic spectra). 
The measured spectra from runs \texttt{theta33} and \texttt{theta43} (green and blue curves in Fig. \ref{fig:thetacompare}(a)) are also similar to a drifting Maxwellian. 
We find that the  maximum electron energy $\gamma_{e,\rm max}$ in runs \texttt{theta33} and \texttt{theta43} is steadily growing over time, indicating sustained Fermi-like acceleration, yet the acceleration rate is much slower than in  quasi-perpendicular shocks. The same is true for runs \texttt{theta13, theta23}, whose results are not shown here. The increase in acceleration rate with magnetic obliquity is primarily driven by the fact that the energy gain per SDA cycle grows monotonically with  $\theta_B$ (see Fig. \ref{fig:SDAtheory}(c)). 

Fermi-like acceleration is expected to be efficient in low-obliquity shocks, albeit with a slow acceleration rate. As demonstrated in Section \ref{sec:waves}, due to the large SDA injection efficiency of quasi-parallel shocks (see Fig. \ref{fig:SDAtheory}(c)), the reflected electrons cause a sufficient temperature anisotropy in the upstream to enable the oblique firehose instability to grow. As Fig. \ref{fig:Bzsigs}(e) (corresponding to $\theta_B=43^\circ$) shows, waves associated with the oblique firehose instability are triggered even with quasi-parallel shocks. By tracing a sample of selected electrons in 
runs \texttt{theta23} and \texttt{theta43}, we have confirmed that a significant fraction of the reflected electrons are undergoing multiple cycles of SDA, by scattering off the upstream waves back toward the shock. This is similar to the behavior 
described in Fig. 8 of Paper I for a quasi-perpendicular shock with $\theta_B=63^\circ$. Thus, we conclude that
Fermi-like acceleration process consisting of multiple cycles of SDA operates efficiently in quasi-parallel shocks, albeit at a slower rate than in quasi-perpendicular shocks. 

In quasi-parallel shocks, we observe that the electron spectra in our simulations peak at a slightly higher kinetic energy ($\gamma_e - 1\sim 0.5$) than expected for the drifting 
Maxwellian at initialization, which peaks at $\gamma_e -1 \sim 0.3$.
In such shocks, we find that a significant fraction of ions propagate upstream, in agreement with the results of hybrid simulations  \citep[e.g.,][]{Caprioli2013,Caprioli2014}. We speculate that the overall heating in the upstream electron spectrum might result from the  interaction with long-wavelength modes driven by the reflected ions.  On top of the ion waves, the electron firehose instability, growing on electron scales, will mediate efficient Fermi-like electron acceleration, as described above.

In summary, the injection process mediated by SDA cannot operate in superluminal shocks. 
In subluminal quasi-perpendicular shocks, the reflection fraction 
is moderate ($\sim 10 - 20\%$) at angles such that $u_t\lesssim v_{{\rm th},e}$. Here, the beam of returning electrons induces a sufficient temperature anisotropy in the upstream such that oblique firehose waves can be generated, mediating long-term Fermi-like electron acceleration via multiple SDA cycles.  Since the average energy gain from each SDA cycle is much larger than the electron thermal energy, the acceleration rate is fast. At obliquity angles such that $u_t\gtrsim v_{{\rm th},e}$, the SDA injection efficiency is poor, resulting in weak temperature anisotropies that do not meet the critical threshold for the excitation of the electron firehose instability. In the absence of firehose-driven waves, the electron acceleration process terminates after one cycle of SDA.
In subluminal quasi-parallel shocks, the SDA efficiency is large, so the upstream waves can be promptly triggered and Fermi-like electron acceleration is very efficient.  However, due to the small energy gain resulting from each cycle of SDA, the acceleration rate is slow.

\subsection{Dependence on the Magnetization $\sigma$}\label{sec:sigma}

\begin{figure}[tbp]
\begin{center}
\includegraphics[height=0.24\textheight]{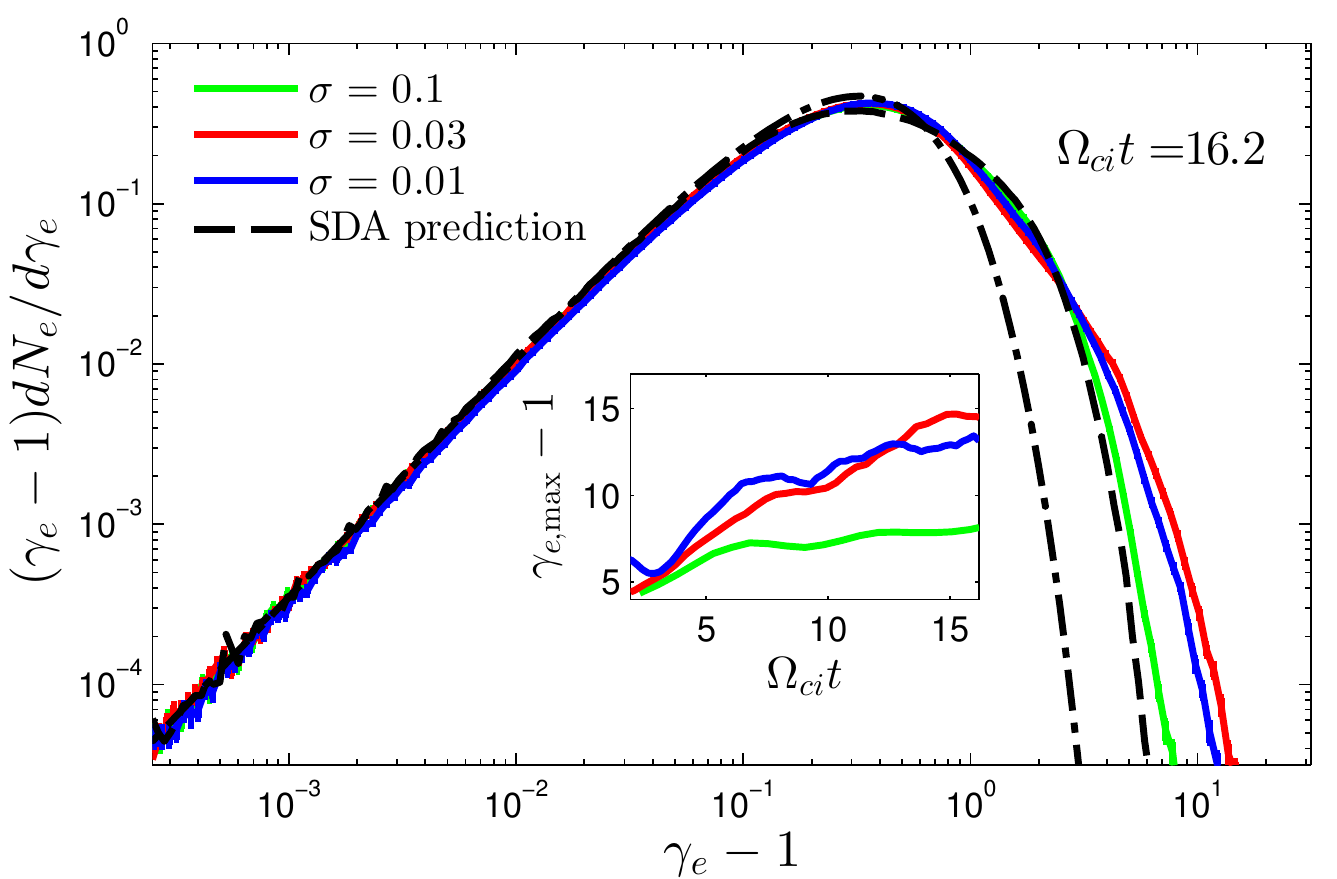}
\caption{Electron upstream energy spectra at $\Omega_{ci}t = 16.2$ from the runs \texttt{sig1e-1\_63, reference} and \texttt{sig1e-2\_63}, as indicated in the legend.
The black dashed curve shows the SDA-synthetic spectrum
for the corresponding pre-shock parameters and the black dot-dashed curve
shows the drifting Maxwellian distribution at initialization, having $u_0 = 0.15 c$ and $T_e = 10^{9}\ K$.
In the run \texttt{sig1e-1\_63}, the 
upstream magnetic pressure is too strong for the electron firehose instability to be triggered, so no upstream waves are generated (Fig. \ref{fig:Bzsigs}) to sustain long term Fermi acceleration, and the maximum energy (red line in the subpanel) stops growing. 
The \texttt{reference} run and \texttt{sig1e-2\_63} show similar results, because the SDA injection is similar, and their low magnetization allows upstream waves to grow (Fig. \ref{fig:Bzsigs}), to mediate long-term Fermi acceleration.}
\label{fig:sigmacompare}
\end{center}
\end{figure}
To explore the effect of the flow magnetization $\sigma$, we have run simulations with $\sigma = 0.1 $, $ 0.01$ and $0.003$, to be compared with our reference case $\sigma=0.03$.
We fix $T_e = 10^9 K$ and $u_0 = 0.15\,c$, so that the Mach number stays fixed at $M_s = 3$. The range from $\sigma = 0.003$ to $\sigma = 0.1$ corresponds to a plasma beta varying from $\beta_p = 200$ to $\beta_p = 6$. At fixed magnetization, we study a few magnetic obliquity angles.

\begin{figure*}[tbp]
\begin{center}
\includegraphics[height = 0.27\textheight]{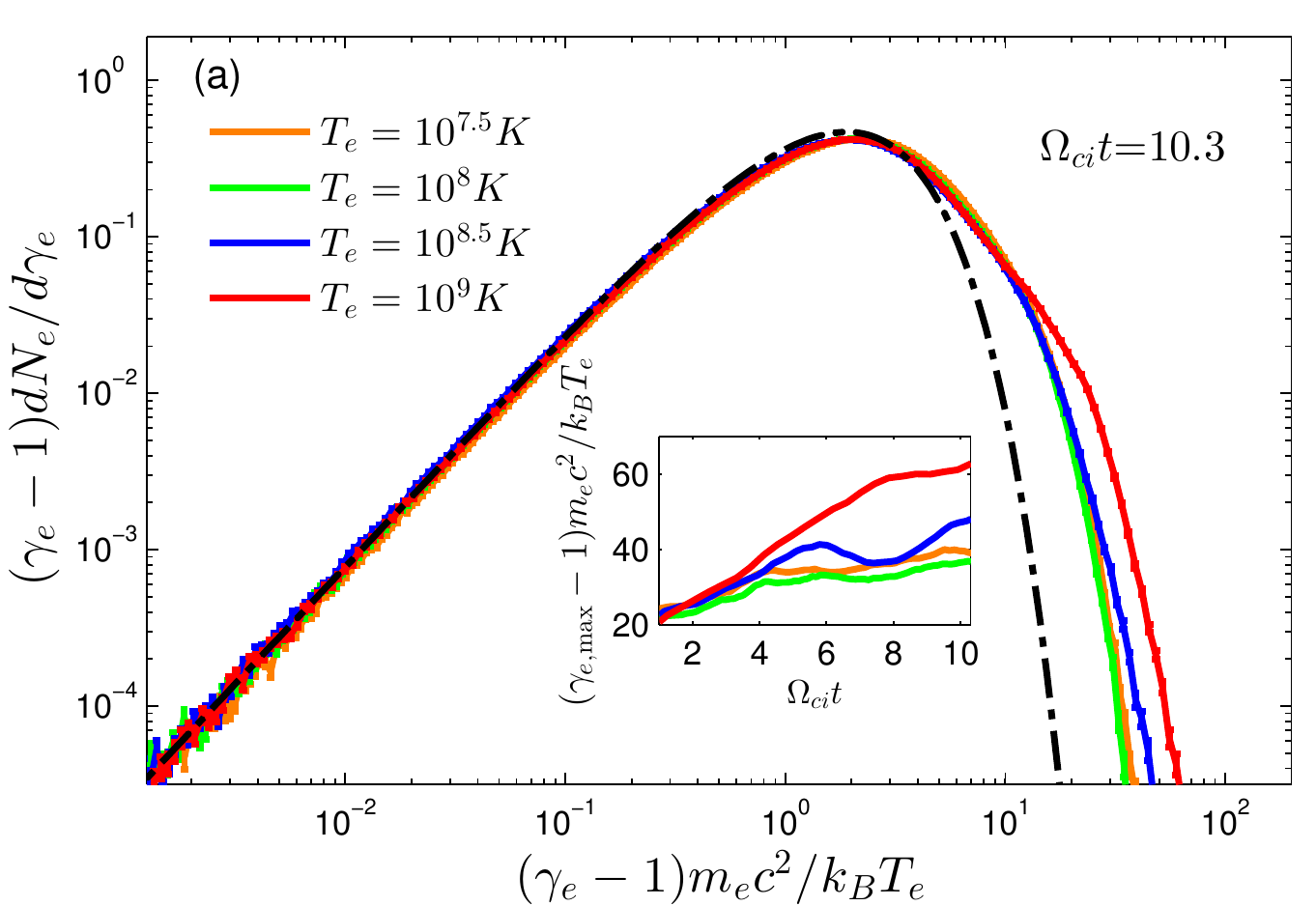}
\includegraphics[height = 0.27\textheight]{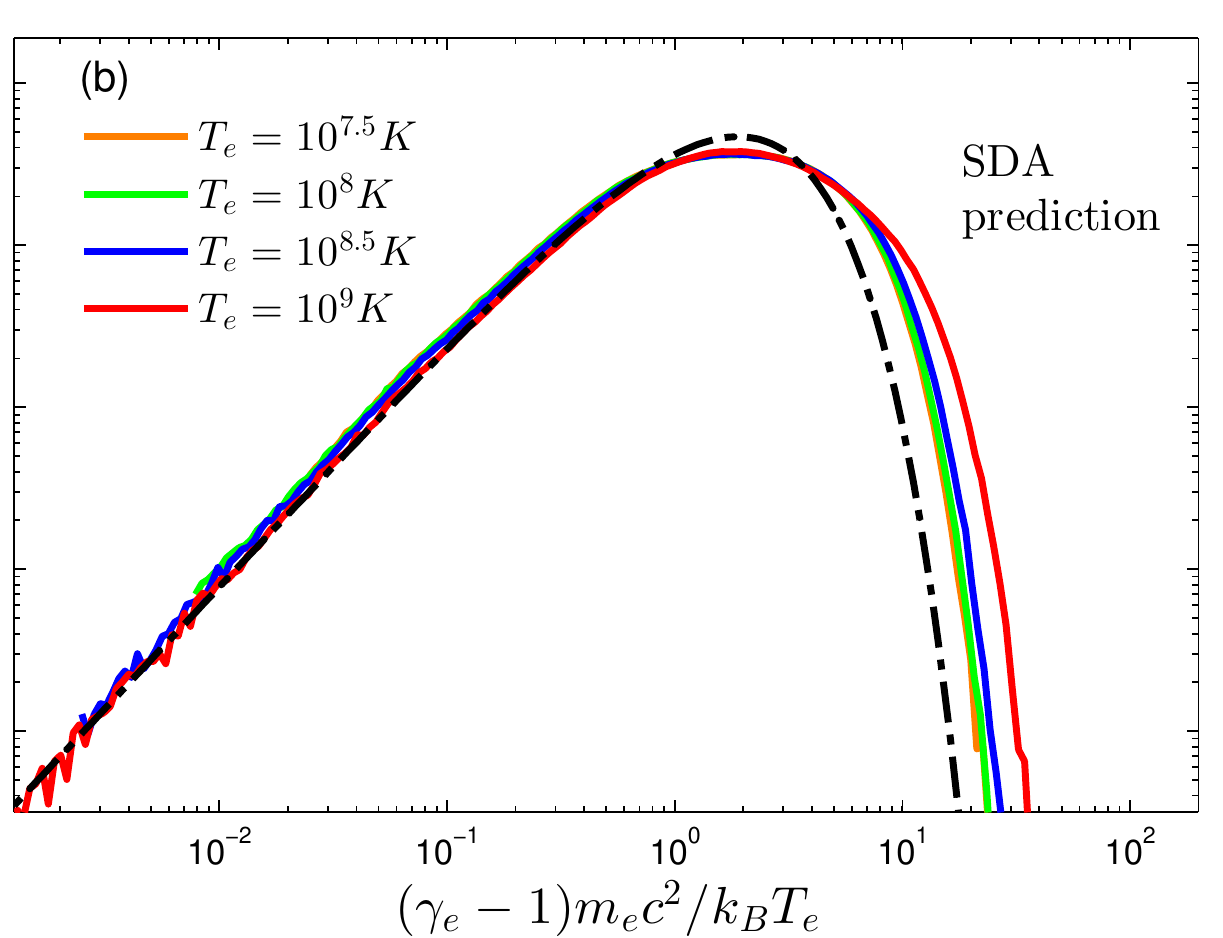}
\caption{Panel (a): upstream electron energy 
spectra measured at $\Omega_{ci}t = 10.3$  from the runs \texttt{Te1e7.5, Te1e8, Te1e8.5, reference}, as indicated in the legend. The subplot shows the temporal evolution of 
the maximum energy of the upstream electrons, in units of $k_B T_e$. 
Panel (b): upstream electron energy spectra predicted by SDA theory. 
In both panels, the black dot-dashed line corresponds to the drifting 
Maxwellian distribution at initialization, having $u_0 = 0.15\,c$ and 
$T_e = 10^9K$.
}
\label{fig:spec_te}
\end{center}
\end{figure*}

We find that the electron acceleration in runs with $\sigma=0.01$ (\texttt{sig1e-2\_43, sig1e-2\_53, sig1e-2\_63, sig1e-2\_73}) shows strong similarities with our reference runs having $\sigma = 0.03$  (\texttt{theta43, theta53, reference, theta73},  respectively). On the other hand, electron acceleration via the Fermi-like process is suppressed at higher magnetizations, in the runs with $\sigma = 0.1$ (\texttt{sig1e-1\_43, sig1e-1\_53, sig1e-1\_63, sig1e-1\_73}).
To illustrate the dependence on $\sigma$, we present in Fig. \ref{fig:sigmacompare} the upstream electron energy
spectra measured at $\Omega_{ci}t = 16.2$ from the runs \texttt{sig1e-2\_63, reference, sig1e-1\_63},
having the same quasi-perpendicular obliquity angle $\theta_B = 63^\circ$. The subplot in Fig. \ref{fig:sigmacompare}
traces the evolution of the maximum energy of the upstream electrons. 
The SDA-synthetic spectrum for the corresponding pre-shock parameters is shown as a black dashed curve in Fig. \ref{fig:sigmacompare}.
We remark that, based on our SDA theory, the SDA-synthetic spectrum has no explicit 
dependence on magnetization. In fact, in the weakly magnetized shocks considered here ($\sigma\ll1$), 
the magnetic field energy does not significantly affect the shock structure, so our assumed values for the magnetic compression ratio $b=4$ and the cross shock potential $\Delta \phi = 0.5$ still apply, regardless of $\sigma$.

As mentioned above, the simulations 
\texttt{sig1e-2\_63} and \texttt{reference} yield similar electron acceleration efficiencies and a comparable acceleration rate (blue and red curves in Fig. \ref{fig:sigmacompare}).
In both cases, $\sim 20\%$ of electrons populate a non-thermal tail 
in the energy spectrum, and the maximum energy evolves steadily to higher and higher values, well beyond the SDA-synthetic spectrum. This suggests that long-term Fermi-like acceleration is operating fast and efficiently.  
On the other hand, the electron acceleration in run \texttt{sig1e-1\_63} does not 
go beyond one cycle of SDA. The maximum energy saturates 
after $\Omega_{ci}t\sim 7$ to a value comparable to the high-energy cutoff in the SDA-synthetic spectrum ($\gamma_e - 1\simeq 7$). 
The shape of the measured electron spectrum (green curve in Fig. \ref{fig:sigmacompare})  resembles closely the SDA-synthetic spectrum. 

The similarity between the runs \texttt{sig1e-2\_63}
and \texttt{reference} stems from the fact that both the injection efficiency and the average energy gain via SDA are nearly identical 
in the two runs, so the returning electrons induce a similar temperature anisotropy in the upstream, irrespective of $\sigma$.
We have demonstrated in Section \ref{sec:waves} that for $\sigma = 0.03$, the  oblique firehose instability is excited across a wide range of magnetic obliquities 
(see Fig. \ref{fig:Bzsigs}(a)-(d)). 
Having comparable electron anisotropy and lower magnetization (so, higher $\beta_e$), the instability threshold is easily met at shocks having $\sigma = 0.01$. The resulting waves mediate efficient  Fermi-like acceleration (the wave patterns of runs \texttt{sig1e-2\_43,sig1e-2\_53,sig1e-2\_63} are shown in Fig. \ref{fig:Bzsigs}(a)-(c), respectively). 
Generalizing this argument to different obliquities, we see why the electron acceleration in runs having $\sigma = 0.01$ (\texttt{sig1e-2\_43, sig1e-2\_53, sig1e-2\_63, sig1e-2\_73}) shows a similar efficiency as in our reference runs with $\sigma = 0.03$, (\texttt{theta43, theta53, reference, theta73},  respectively). 
In contrast, in shocks with  $\sigma = 0.1$ (so, lower $\beta_e$), the electron anisotropy is not sufficient to satisfy the threshold criterion in Equation \eqref{eq:threshvalue}, so the growth of oblique firehose waves is inhibited  (see Fig. \ref{fig:Bzsigs}(i)-(l)). In the absence of upstream waves, the electron acceleration process stops after one cycle of SDA. 

In summary, the dependence of the electron acceleration physics on magnetization
in weakly to moderately magnetized shocks ($\beta_p \gtrsim $ a few) is largely determined by whether the upstream magnetic pressure is weak enough to allow the growth of the electron firehose instability, thereby allowing the electrons to participate in long-term Fermi-like acceleration process. 
The instability is suppressed in shocks having $\sigma = 0.1\ (\beta_p = 6)$, the strongest 
magnetization we have explored. Here, the electron acceleration process stops after one cycle of SDA. On the other hand, 
lower magnetizations ($\sigma = 0.01 - 0.03$, corresponding to $\beta_p = 20 - 60$) allow firehose-driven waves to grow and support the Fermi-like process. We argue that this mechanism can operate down to at least $\sigma = 0.003\ (\beta_p = 200)$, 
since we find that in run \texttt{sig3e-3\_63} the same type of waves  are generated in the upstream region. 
At lower magnetizations (or equivalently, higher $\beta_p$), the ordinary-mode instability is likely to dominate \citep{Lazar2009,Lazar2010,Lazar2014}, in analogy to unmagnetized shocks. A discussion of the electron injection and acceleration mechanism at extremely low $\sigma$ (high $\beta_p$), is likely to differ from the scenario presented here, and is beyond the scope of this paper.

\subsection{Dependence on the Electron Temperature $T_e$ }\label{sec:te}

\begin{figure}[tbp]
\begin{center}
\includegraphics[width = 0.8\textwidth]{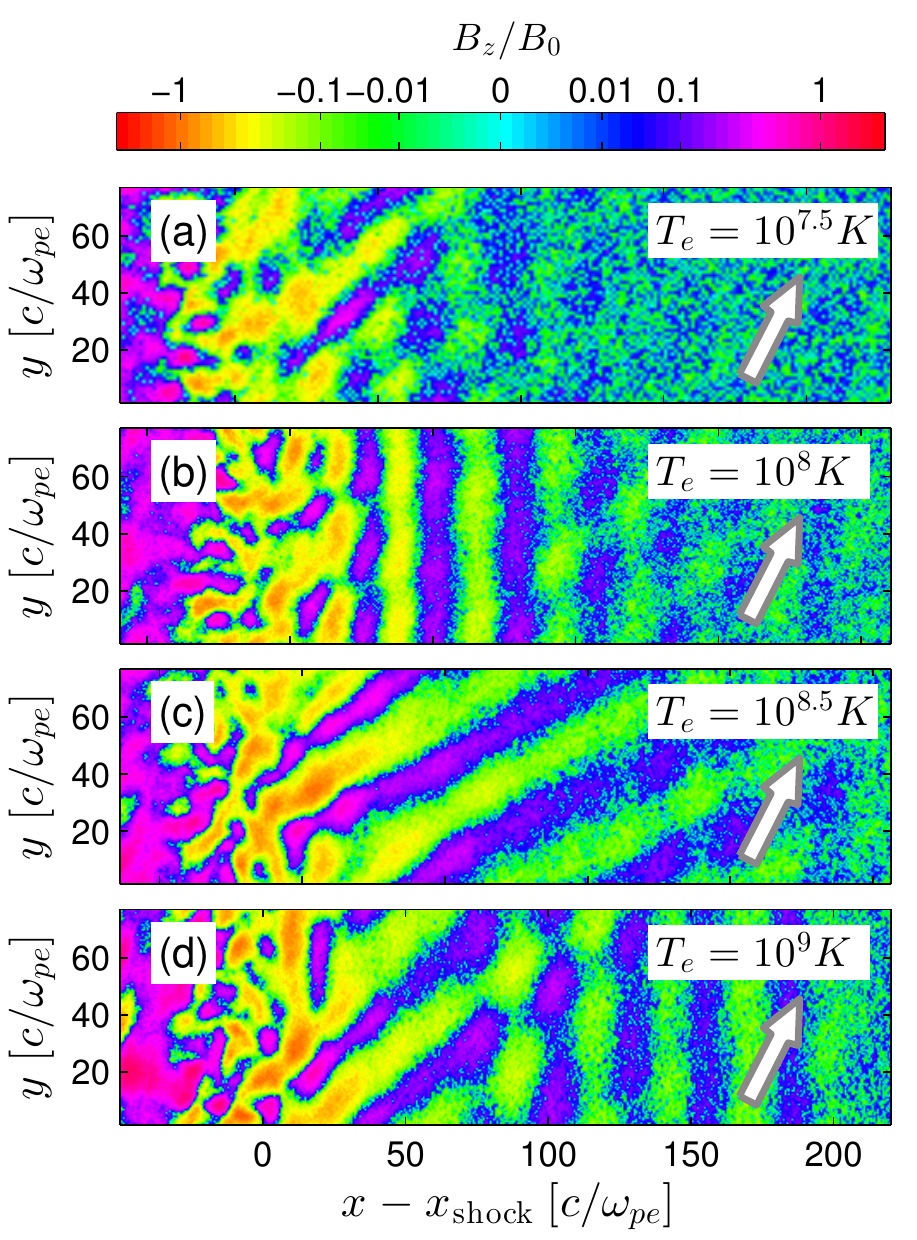}
\caption{2D plots of $B_z/B_0$ at $\Omega_{ci}t=10$ from the runs \texttt{Te1e7.5, Te1e8, Te1e8.5, reference}.
The orientation of the upstream magnetic field is indicated by the white arrows. }
\label{fig:Bztes}
\end{center}
\end{figure}
To investigate the dependence of 
electron acceleration on the upstream electron temperature $T_e$, 
we perform simulations with lower temperatures than the \texttt{reference} run (which has $T_e = 10^{9}K$). We vary the temperature  from $10^{7.5}K$ to $10^{8.5}K$ at fixed $\theta_B = 63^\circ$, $\sigma = 0.03\ (\beta_p = 20)$ and $M=2\ (M_s = 3)$. In order to keep the Mach number $M$ fixed, we scale the upstream bulk flow velocity  as $u_0$ $\propto\sqrt{T_e}$ (Table \ref{table:params}). 

Fig. \ref{fig:spec_te}(a) shows the upstream electron energy spectra at $\Omega_{ci}t= 10.3$ from runs \texttt{Te1e7.5, Te1e8, Te1e8,5, reference},  while Fig. \ref{fig:spec_te}(b) shows the corresponding SDA-synthetic spectra.
Unlike the other electron spectra presented in this section, 
the horizontal axis of these spectra measures $(\gamma_e - 1)m_e c^2/k_B T_e$, instead of 
$\gamma_e -1$. This choice is motivated by the fact that the electron spectra from runs with different $T_e$ peak at 
$\gamma_e - 1 \sim k_B T_e$, so comparisons are easier after rescaling with $k_B T_e$. In addition, 
we have shown in Section \ref{sec:SDA} that the average energy gain in one SDA cycle also scales with $T_e$ (Fig. \ref{fig:SDAtheory}(d)), which further motivates our rescaling.

We find that in all the runs considered here (\texttt{Te1e7.5, Te1e8, Te1e8,5, reference}), the maximum electron energy $\gamma_{e,\rm max}$ evolves well beyond the prediction of a single cycle of SDA (see subpanel of Fig. \ref{fig:spec_te}(a)), indicating that long-term Fermi acceleration process operates efficiently in all the runs. 
This stems from the fact that the electron temperature anisotropy is only weakly dependent on $T_e$ (see Fig. \ref{fig:wavegrowth}), so the threshold for the oblique firehose instability is still met, and the upstream waves can grow and mediate long-term Fermi-like acceleration. 
Indeed, we observe similar wave patterns in all the runs mentioned above, regardless of $T_e$ (see Fig. \ref{fig:Bztes}).\footnote{The waves in the run \texttt{Te1e7.5} look weaker, due to limited numerical accuracy at low temperatures. The strength of electromagnetic fields is weaker in lower temperature runs (e.g., $B_0\propto u_0\propto \sqrt{T_e}$), resulting in a lower signal to noise ratio.}

As regards spectra, we find that, when the kinetic energy $(\gamma_e - 1)m_e c^2$ is measured in units of $k_BT_e$, the
electron energy spectra from runs \texttt{Te1e7.5} and \texttt{Te1e8} nearly overlap. Both the normalization and the maximum energy of the non-thermal tail are almost identical. The agreement between these two runs can be understood from the SDA theory, since 
both the reflection fraction and the average energy gain (in units of $k_B T_e$) stay almost constant in the regime 
$T_e \lesssim 10^8 K$ (see Fig. \ref{fig:SDAtheory}(d)).
With increasing temperature towards the relativistic regime (runs \texttt{Te1e8.5} and \texttt{reference}), the normalization of the non-thermal tail tends to decrease, but the spectrum extends to higher energies (see Fig. \ref{fig:spec_te}(a)). Once again, SDA theory predicts both the lower injection efficiency and the higher maximum energy, as the temperature increases beyond $T_e \gtrsim 10^8K$ (see Fig. \ref{fig:SDAtheory}(d)).

To summarize, Fermi-like electron acceleration operates efficiently over the whole  temperature range we have explored, $T_e = 10^{7.5}K - 10^{9}K$, with fixed $M_s = 3$, $\sigma = 0.03$ and $\theta_B = 63^\circ$. 
The number density of non-thermal electrons stays roughly constant in the regime $T_e \lesssim 10^8 K$, but decreases
slightly  with increasing $T_e$ for $T_e \gtrsim 10^8 K$, as suggested by
Fig. \ref{fig:SDAtheory}(d). The acceleration rate is 
faster in the higher temperature regime $T_e \gtrsim 10^8 K$, since  the average energy gain per SDA cycle increases with $T_e$. The acceleration rate at lower temperatures  is slower, but it saturates at a constant value for $T_e \lesssim 10^8 K$.
Based on the SDA theory presented in Section \ref{sec:SDA} and on the mechanism of wave generation described in Section \ref{sec:waves}, we expect that our results can be extrapolated to even lower temperatures (at fixed Mach number and magnetic obliquity) since both the SDA injection efficiency and the threshold for excitation of the firehose instability are independent of $T_e$.

\subsection{Dependence on the Mach number $M_s$}
We comment on the dependence of the electron acceleration physics 
on the Mach number $M_s$, based on simulations presented in \cite{Narayan2012}.

From the discussion above, we know that the electron acceleration efficiency is ultimately related to the efficiency of SDA injection. In turn, the number of SDA-reflected electrons and their anisotropy determine whether the oblique firehose instability can grow in the upstream, governing the long-term Fermi acceleration.
A key parameter that regulates the SDA injection efficiency is the HT velocity $u_t$, which scales linearly with $M_s$ (Equation \eqref{eq:ut}). 
When $u_t \gtrsim v_{{\rm th},e}$, or equivalently $M_s\sec\theta_B\sqrt{m_e/m_i}\gtrsim 1$, SDA injection is expected to be inefficient.
For $\theta_B \gtrsim 45^\circ$
and mass ratio $m_i / m_e = 100$ (as employed in our reference runs), the requirement $u_t\lesssim v_{{\rm th},e}$ for efficient SDA injection is satisfied for Mach numbers $M_s \lesssim5.5$, 
while for the realistic mass ratio $m_i/m_e = 1836$ the requirement is $M_s \lesssim 23$. As the Mach number increases towards this limit, the acceleration efficiency is expected to decrease. In contrast, since the energy gain per SDA cycle increases with $u_t\propto M_s$, the acceleration rate will be faster for higher $M_s$. 

The results of 2D PIC simulations presented in Fig.2(b) of \cite{Narayan2012} illustrate the dependence on $M_s$. 
There, the Alfv\'enic Mach number was fixed at $M_A = 8$, which corresponds 
to $\sigma = 0.03$, as in our \texttt{reference} run. 
The electron temperature is changed but the upstream flow speed is fixed, which 
effectively results in varying the Mach number. For the temperature range $T_e = 5\times 10^7 K-10^9 K$ explored in \cite{Narayan2012}, the Mach number varies between $M_s\simeq2$ and $9$ ($M_s \propto T_e^{-1/2}$). The electron spectra from 
runs of different $M_s$ show that the normalization of the non-thermal tail decreases monotonically with increasing $M_s$. 
In particular, the acceleration efficiency drops from $\sim 10\%$ in the run with $M_s = 2$ $(T_e = 10^9\ K)$ to $\sim 4\%$ in the run with $M_s \simeq 4.5$ $(T_e = 2\times 10^8\ K)$, and it is negligible ($\ll 1\%$) in the runs having $M_s\gtrsim 6$ $\ (T_e\lesssim 10^8 K)$. These results are in agreement with our arguments above. 

For higher Mach number shocks, 
a regime relevant for supernova remnants,
 injection via SDA becomes extremely inefficient 
and other pre-acceleration mechanisms, such as 
shock surfing acceleration or injection via whistler waves, will 
dominate \citep[see, e.g.,][]{Dieckmann2000,Hoshino2002,Schmitz2002,Amano2007,Riquelme2011,Matsumoto2012}.

\section{Summary and Discussion}\label{sec:discussion}
In this paper, the second of a series, we complete our investigation of electron acceleration in 
low Mach number shocks ($M_s = 3$) by performing a suite of self-consistent
2D PIC simulations. 
In Paper I, we studied a reference shock that propagates in a high-temperature plasma ($T_e = 10^9 K$) carrying a quasi-perpendicular magnetic field (with magnetization $\sigma=0.03$ and obliquity $\theta_B=63^\circ$). We identified a Fermi-like electron acceleration mechanism whose injection is governed by shock drift acceleration (SDA).
A fraction of the incoming thermal electrons are 
reflected at the shock front by the mirror force of the shock-compressed field, and they
are energized by the motional electric field while drifting along the shock surface.
The reflected electrons propagate ahead of the shock, where their interaction with the upstream flow generates oblique magnetic waves in the upstream region. The waves 
 scatter the reflected electrons back towards
the shock for multiple cycles of SDA, in a process resembling the Fermi mechanism.

In the present work, we address the nature of the upstream waves, which are 
essential for maintaining the long-term Fermi-like acceleration. 
Using 2D periodic box simulations in the upstream frame, we study the interaction between the beam of SDA-reflected electrons and the pre-shock plasma. We 
 confirm that the upstream waves are triggered by the electrons reflected at the shock during the SDA process. The distribution of reflected electrons is anisotropic, such that the temperature parallel to the field is larger than  perpendicular ($T_{e\parallel}>T_{e\perp}$).
We demonstrate that the waves are associated with the oblique mode of the electron firehose instability, 
which is driven by the electron temperature anisotropy and requires the pressure associated to the electron anisotropy
to be stronger than the plasma magnetic pressure. It follows that the waves cannot be generated if the fraction of SDA-reflected electrons is too small or if the upstream magnetic field is too strong (i.e., low beta plasmas). In the absence of upstream magnetic waves, the Fermi-like acceleration process will be inhibited.

By means of fully-consistent 2D shock simulations, we systematically explore the dependence of the electron acceleration efficiency on the pre-shock conditions for low Mach number shocks ($M_s = 3$).
We investigate the effect of the upstream magnetic field obliquity $\theta_B$, of the magnetization $\sigma$ and of the electron temperature $T_e$.
 
We find that at superluminal shocks (i.e., where the de Hoffman-Teller velocity exceeds the speed of light), the SDA process does not operate and no electron is accelerated.
In subluminal shocks, the efficiency of electron acceleration depends on the magnetic obliquity $\theta_B$. Injection via SDA into the acceleration process is inefficient if $u_t\gtrsim v_{{\rm th},e}$, where $v_{{\rm th},e}$ is the electron thermal speed and the de Hoffman-Teller velocity $u_t$ can be written as $u_t\sim v_{{\rm th},e}M_s\sec\theta_B\sqrt{m_e/m_i}$. Here, few electrons are able to propagate back into the upstream, so the resulting temperature anisotropy induced in the pre-shock region is too weak to trigger the firehose instability. In the absence of upstream waves, the process of electron acceleration stops after one cycle of SDA. In contrast, for $u_t\lesssim v_{{\rm th},e}$ (still, for quasi-perpendicular shocks with $\theta_B\gtrsim45^\circ$), the electron acceleration process is efficient and fast. In fact, the fraction of SDA-reflected electrons is large enough to generate firehose-driven waves in the upstream. These waves help the Fermi-like process by enabling multiple SDA cycles. Also, the acceleration rate is fast because each SDA cycle provides a significant energy gain. The Fermi-like process operates efficiently also in quasi-parallel shocks ($\theta_B\lesssim45^\circ$), but the electron acceleration rate is slower. In fact, the SDA reflection efficiency increases at lower $\theta_B$, but the energy gain per SDA cycle is smaller. In addition, in quasi-parallel shocks, we find that a fraction of the incoming ions are reflected back into the upstream. The ion acceleration physics operates on timescales longer than the timespan of our simulations, but it is not expected to modify significantly the process of firehose-mediated electron acceleration described above.

When varying the magnetization $\sigma$ at fixed $T_e = 10^9\ K$, we find that the electron acceleration physics does not depend on the flow magnetization so long as $\sigma\lesssim 0.03$ . Neither the injection efficiency of SDA nor the energy gain per SDA cycle depends explicitly on $\sigma$, and the threshold condition for the excitation of the oblique firehose mode is satisfied at all $\sigma\lesssim 0.03$, as long as the magnetic obliquity is such that $u_t\lesssim v_{{\rm th},e}$. Oblique firehose waves are still present at $\sigma=0.003$, but we expect that the injection and acceleration physics at yet lower magnetizations could be different, as the shock transitions to the Weibel-mediated regime relevant for unmagnetized flows. 
In contrast, at high magnetizations ($\sigma = 0.1$), even though injection via SDA is efficient, the Fermi process cannot operate because the strong magnetic pressure in the upstream suppresses the growth of the  firehose instability. 

When varying the electron temperature $T_e$ at fixed $\sigma = 0.03$ and $\theta_B = 63^\circ$, 
we find efficient long-term electron acceleration
across the whole temperature range $T_e=10^{7.5}\ K - 10^9\ K$. Both the SDA injection efficiency and the acceleration rate are insensitive to the electron temperature in the regime of non-relativistic electrons ($T_e\lesssim 10^8\ K$). No major change is observed for trans-relativistic temperatures ($T_e\gtrsim 10^{8.5}\ K$), except that there is a slight tendency for a higher SDA injection efficiency and  a larger energy gain per SDA cycle, as $T_e$ increases.

In summary, our study finds that efficient Fermi-like electron 
acceleration, whose injection is controlled by SDA,
operates in low Mach number shocks 
for a variety of pre-shock conditions. The Fermi acceleration is mediated by oblique upstream waves generated by the electron firehose instability, which can only be excited if the plasma beta in the upstream region is sufficiently large. 
As the 
criterion in Equation \eqref{eq:threshvalue} suggests,
the electron firehose instability would be completely suppressed 
for $\beta_{e\parallel}\lesssim 1.3$, which 
corresponds to $\beta_p \lesssim 2.6$. 
For $\beta_p= 20-200$, we have demonstrated that
the growth of the upstream firehose waves is allowed, given a sufficient electron anisotropy. In this respect, our work is complementary to earlier PIC studies of shocks propagating in low-beta plasmas (so, with high Mach number), as appropriate for supernova remnants. There, firehose-driven waves cannot grow, since $\beta_p \lesssim 1$. Also, a different injection mechanism, other than SDA, is required for efficient electron acceleration.
For instance, \cite{Riquelme2011}
found that in shocks with higher sonic Mach number ($M_s\simeq 7$, as opposed to our choice $M_s=3$) and lower plasma beta ($\beta_p \lesssim 1$), electron injection is regulated by the interaction with oblique whistler waves near the shock front. Alternatively, \cite{Matsumoto2012} found that the shock surfing mechanism
serves to inject electrons into Fermi acceleration at shocks with low plasma beta ($\beta_p \ll 1$) and high Alfv\'enic Mach number ($M_A\sim 30$) \citep[see also e.g.][]{McClements2001,Hoshino2002,Amano2007}.

The generality of our Fermi-like 
electron acceleration mechanism in low Mach number shocks 
offers a possible solution to the problem of electron injection in 
 merger shocks of  galaxy clusters. 
The bright radio emission that is observed from radio relics cannot be reconciled with the  poor efficiency of the commonly-invoked ``thermal leakage'' model for electron injection  \citep{Malkov1998,Gieseler2000,Kang2002}.
As we discussed in Paper I, the thermal leakage model assumes that the electrons are scattered by downstream magnetic waves back into the upstream. This requires that 
the electrons have large momentum, a few times larger than the characteristic post-shock ion momentum,  so that their Larmor radius is larger than the scale of the magnetic turbulence. The number of incoming thermal electrons that satisfy this stringent criterion is extremely small.
In contrast, our mechanism, based on first-principles PIC simulations, does not involve 
any scattering by the downstream turbulence. Rather,  
the shock itself acts as a magnetic mirror, reflecting a fraction of the incoming electrons back upstream via SDA.  
The minimum electron momentum required for reflection via SDA is much lower (by a factor of $\sim m_e/m_i$) than that required in the thermal leakage model. For this reason, the electron injection fraction in our low Mach number shocks is as large as $\sim 10\%-20\%$, which can explain the bright radio emission of galaxy cluster shocks.\footnote{Incidentally, electrons accelerated in radio relics are also invoked as a seed population of relativistic electrons for particle reacceleration by turbulence at radio halos \citep{Brunetti2011}.}

In the thermal leakage model, due to the stringent constraint on the minimum momentum for electron injection, 
the number of accelerated ions is expected to exceed 
that of accelerated electrons by a large factor.
The high-energy ions will interact with the thermal gas in the ICM 
and produce gamma-ray emission. 
Assuming the large ratio of ion-to-electron acceleration efficiencies predicted by the thermal leakage model, 
\cite{Vazza2014} found that, 
given the current observations of radio relics, which are 
powered by synchrotron emission of the shock-accelerated electrons, the predicted gamma-ray luminosity of nearby galaxy clusters, resulting from the accelerated ions, should be above the detection limit of the 
\textit{Fermi} telescope. 
Yet, \textit{Fermi} has not detected any gamma-ray signature from these
systems. This apparent tension can be alleviated if 
the electron acceleration efficiency is much higher than expected from the thermal leakage model, as indeed predicted by our mechanism.\footnote{The possibility of a higher ratio of electron-to-ion acceleration efficiencies  -- as a solution to the lack of {\it Fermi} detections of galaxy clusters --
 has already been  invoked by \cite{Brunetti2014}.}  In particular, we find that a fraction as large as $\sim 10\%-20\%$ of electrons can be accelerated in quasi-perpendicular shocks, where ion acceleration is known to be extremely inefficient  \citep[e.g.,][]{Caprioli2013,Caprioli2014}.
The ratio of electron-to-ion acceleration efficiency should then be higher than expected from the thermal leakage model, suggesting that the gamma-ray brightness of galaxy cluster shocks is likely to be significantly lower than estimated by \cite{Vazza2014}. This could explain the lack of  \textit{Fermi} detections of galaxy clusters.

Our study might also help to clarify why some low Mach number shocks are not efficient electron accelerators.
Using \textit{Chandra} X-ray images, \cite{Russell2011} 
have unambiguously identified two merger shocks with $M_s \simeq 2.1$ and $M_s \simeq 1.6$
in the  galaxy cluster Abell 2146. However, 
no radio emission is detected there. Currently, no convincing explanation has been proposed.
We point out that the merger shocks in 
Abell 2146 are not located 
at the outskirts of the galaxy cluster, 
 where most radio relics have been detected. 
Near the cluster center, the magnetic 
field should be stronger than in the outskirts, though 
no measurement of magnetic field strength is available  at the location of these two shocks. 
The pre-shock temperature and density inferred from X-ray observations \citep{Russell2010} suggest that 
$T_e \sim 6\times 10^7\ K$ and $n_e \sim 10^{-3}$ at both shocks. 
If $B_0\sim 8\,\mu{\rm G}$ (slightly stronger than typically field strength inferred in cluster outskirts), 
the plasma beta could be as low as $\beta_p \sim 2.5$, which would prevent the 
growth of the electron firehose instability. 
Without upstream waves, the 
process of electron acceleration would stop after one cycle of SDA, and electrons would not be accelerated to relativistic energies.
This argument may also offer a generic 
explanation for the rarity of radio relics in the central regions of galaxy clusters, as discussed 
by \cite{Vazza2012}.

\section*{Acknowledgements}
X.G. and R.N. are supported in part by NASA grant NNX14AB47G.
X.G. thanks Philip Mocz for helpful comments on the manuscript and Pierre Christian for useful discussions. 
We thank R. van Weeren, P. Nulsen, G. Brunetti, F. Vazza, and the anonymous referee for helpful comments. 
L.S. is supported by NASA through Einstein
Postdoctoral Fellowship grant number PF1-120090 awarded by the Chandra
X-ray Center, which is operated by the Smithsonian Astrophysical
Observatory for NASA under contract NAS8-03060. 
The computations in this paper were run on the Odyssey cluster supported by the FAS Division of Science, Research Computing Group at Harvard University, on XSEDE resources under contract No.TG-AST120010, and on NASA High-End Computing
(HEC) resources through the NASA Advanced Super-
computing (NAS) Division at Ames Research Center.

\bibliographystyle{apj}
\bibliography{accel}

\end{document}